\documentclass[aps,pre,showpacs,showkeys,floatfix,superscriptaddress]{revtex4-1}
\usepackage{amssymb}
\usepackage{amsmath}
\usepackage{amscd}
\usepackage{graphicx}
\usepackage{amsbsy}
\usepackage{color}
\usepackage{bm}
\begin{document}

\title{Self-assembly of hard helices: a rich and unconventional polymorphism}

\author{Hima Bindu Kolli}
\affiliation{Dipartimento di Scienze Molecolari e Nanosistemi, Universit\`{a} Ca' Foscari di Venezia,
Dorsoduro 2137, 30123 Venezia, Italy}
\author{Elisa Frezza}
\affiliation{Dipartimento di Scienze Chimiche, Universit\`{a} di Padova, 
via F. Marzolo 1, 35131 Padova, Italy. Present address:Bases Moléculaires et Structurales des Syst\'emes Infectieux, Uni. Lyon I/CNRS UMR 5086,IBCP,
7 Passage du Vercours, 69367 Lyon, France} 
\author{Giorgio Cinacchi}
\affiliation{Departamento de F\'{i}sica Te\'{o}rica de la Materia Condensada and
Instituto de F\'{i}sica de la Materia Condensada, Universidad Aut\'{o}noma de Madrid,
Campus de Cantoblanco, 28049 Madrid, Spain}
\author{Alberta Ferrarini}
\affiliation{Dipartimento di Scienze Chimiche, Universit\`{a} di Padova, 
via F. Marzolo 1, 35131 Padova, Italy}
\author{Achille Giacometti}
\affiliation{Dipartimento di Scienze Molecolari e Nanosistemi, Universit\`{a} Ca' Foscari di Venezia,
Dorsoduro 2137, 30123 Venezia, Italy}
\author{Toby S. Hudson}
\affiliation{School of Chemistry, University of Sydney, NSW 2006, Australia}
\author{Cristiano de Michele}
\affiliation{Dipartimento di Fisica, Universit\`{a} di Roma "La Sapienza",
Piazzale A. Moro 5, 00185 Roma, Italy}
\author{Francesco Sciortino}
\affiliation{Dipartimento di Fisica, Universit\`{a} di Roma "La Sapienza",
Piazzale A. Moro 5, 00185 Roma, Italy}

\date{today}

\begin{abstract}
Hard helices can be regarded as a paradigmatic elementary model for 
a number of natural and synthetic soft matter systems, 
all featuring the helix as their basic structural unit:
from natural polynucleotides and polypeptides to synthetic helical polymers;
from bacterial flagella to  colloidal helices.
Here we present an extensive investigation of 
the phase diagram of hard helices using a variety of methods.
Isobaric Monte Carlo numerical simulations are used to trace the phase diagram:
on going from the low-density isotropic to the high-density compact phases,
a rich polymorphism is observed exhibiting 
a special chiral screw-like nematic phase and a number of chiral and/or polar smectic phases.  
We present a full characterization of the latter, 
showing that they have unconventional features, 
ascribable to the helical shape of the constituent particles.  
Equal area construction is used to locate  
the isotropic--to--nematic phase transition, and results
are compared with those stemming from an Onsager-like theory.
Density functional theory is also used to study the  
nematic--to--screw-nematic phase transition: 
within the simplifying assumption of perfectly parallel helices, 
we compare different levels of approximation, 
that is second- and third-virial expansions and Parsons-Lee correction.
\end{abstract}
\maketitle
\section{Introduction}
Short-range repulsive interactions are those mainly responsible for 
the structure of classical particle fluid systems; 
this is what originally conferred worthiness to hard--body particle models.\cite{Barker76}
These have actually proven to be a very good representation of 
colloidal particle systems, 
with a very good agreement between   
the theoretical phase diagram of hard spheres and 
the experimental phase behaviour of colloidal spheres.\cite{Pusey86}
Today, ever-increasing importance of colloids and 
advances in the synthesis of colloidal particles of non-spherical symmetry 
\cite{Glotzer07,Sacanna13a,Sacanna13b}
are depriving theoretical studies of any hard particle system of 
what is remaining of a mere academic aura,
demonstrating that these studies may be crucial for 
the design of new colloidal materials.\cite{Solomon11} 

Helical particles are especially worth investigating as 
Nature has conferred to the helix a rather prominent role.
Helical polynucleotides and polypeptides function at 
large enough densities that 
the details of their shape start to be relevant.\cite{Stevens14}
The desire of mimicking Nature in reproducing 
the functions carried out by helical biopolymers has, in turn, led to 
a very active area in polymer research --
the synthesis and characterisation of helical polymers,
aiming at exploiting the inherent chirality of
the helical structure to produce new functional materials
to be used especially in asymmetric catalysis and enantiomeric separation. 
\cite{Nakano01,Yashima09} 

This material interest merges with its inherent  biological interest
in the currently pursued attempt to employ DNA, 
perhaps the most emblematic of all helical biomolecular systems, 
as a building-block for new materials.\cite{Seeman03,Douglas09}

Rather surprisingly, in spite of this wealth of inspiration sources,
helices appear to have been mostly overlooked in 
past theoretical studies on hard-body non-spherical particle systems,
focussing mostly on rod- or disc-like particles,\cite{Allen93,Tarazona08}
possibly due to the tacit assumption that 
helices, as elongated objects, can be assimilated to rods.

To fill this gap, 
we have undertaken a systematic investigation of 
the phase behaviour of hard helices,
using numerical simulation and density functional theory. 
We have found
how finely the isotropic--nematic phase boundaries depend on 
the structural parameters defining a helical particle, 
with a dependence not simply rooted in its aspect ratio, 
thus making a mapping onto an effective rod rather loose.\cite{Frezza13} 
More importantly, we have also provided evidence for
the existence of a new chiral nematic phase, 
named screw-nematic, 
the helix twofold symmetry axes spiral around the main phase director.\cite{Kolli14} 
This was the phase observed in experiments on systems of 
colloidal helical filaments,\cite{Barry06} 
but our results on such a basic model
suggest 
this screw-nematic phase to be  
a general feature of any helical particle system, including DNA suspensions at sufficiently high densities.\cite{Manna07}

In the present work, we build upon past work by extending it in several respects.
(i) We present a complete phase diagram in the density-pressure plane, 
with a special emphasis on 
the smectic phases occurring at densities higher than 
those typical of the conventional and screw-nematic phases, 
and discuss how the screw-nematic order merges with 
a tendency to layering to produce
new chiral, screw-like, smectic phases. 
(ii) We perform a detailed study of the isotropic--nematic coexistence;
(iii) We extend the second-virial theory 
for the nematic--to--screw-nematic phase transition  \cite{Kolli14}
by adding the third-virial contribution, 
and validate it against numerical simulations.

In the next section, 
we provide details on the various theoretical and computational methods used.
We first describe the Monte Carlo simulation technique and then 
the density functional theory at different levels of approximation.
Section III presents and discusses the results, sub-divided in several parts.
In the first, phase diagrams, 
as obtained from isobaric Monte Carlo simulations, are shown and 
the structure of the smectic phases occurring in 
the higher density regions is described by means of 
positional and orientational order parameters and pair correlation functions. 
In the second part, attention is paid to
the isotropic--nematic phase transition,
with the aim of a proper location of the coexisting densities and pressure.  
The third part presents the theoretical results for 
the nematic--to--screw-nematic phase transition. 
Finally, Section IV concludes this work by giving a brief summary and 
possible outlooks.

\section{Model and Methods}
\begin{figure}[htbp]
\begin{center}
   \includegraphics[width=2cm]{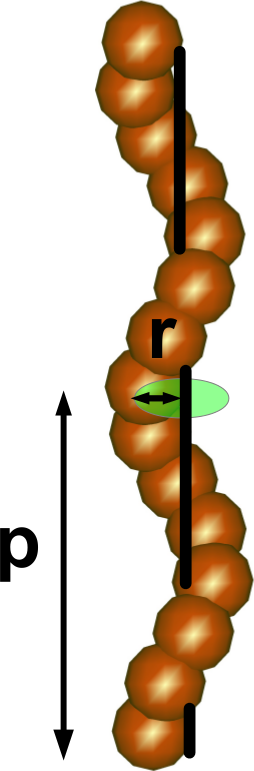}
\end{center}  
\caption{Model for a hard helix, with $r$ its radius and $p$  the pitch. 
}
\label{fig:fig1}
\end{figure}

Helices were simply modelled as 
a line of 15 fused hard spherical beads of diameter $D$ 
rigidly arranged into a helicoidal shape, 
with a contour length fixed to $L=10 D$.\cite{Frezza13}
On changing the radius $r$ and pitch $p$ at fixed $L$, 
the shape of the helical particle can be tuned  
from a straight rod to very wound coils  (Fig.\ref{fig:fig1}). 
We focus on increasingly twisted helical shapes 
in the range  $r/D \in [0.2;0.4]$ and $p/D \in [2;8]$. 
Such particles have a sufficiently large effective aspect ratio 
to display a rich polymorphic liquid-crystal phase behaviour, 
and yet they have an  
intermediate degree of "curliness" ($L/p = {\cal O}(1)$), 
so that phase sequences and phase structures are expected 
to depend sensitively on the overall set of parameters defining the particle shape. 
Hereafter, all lengths 
will be expressed in units of $D$.

To trace the full phase diagrams of such objects, 
we resorted to Monte Carlo (MC) numerical simulations in 
the isobaric(-isothermal) ensemble  (MC-NPT)  \cite{Allen87,Frenkel02}.  
These calculations were preceded by
the construction of the initial compact configurations.
Additional MC simulations in the canonical ensemble (MC-NVT) \cite{Allen87,Frenkel02} 
were performed in one specific case
to identify the precise value of the 
volume fractions 
and pressure 
at the coexistence.
Onsager theory, in different forms,\cite{Onsager49,Parsons79,Lee87} was used for
studying the isotropic-to-nematic and nematic--screw-nematic 
phase transitions.
The remainder of this section provides details on the various
methods used to investigate the phase behaviour.
\subsection{Isopointal search method (ISM)}
One of the difficulties arising in simulations of non-spherical objects
stems from the choice of a judicious set of initial conditions that allow a correct span of the whole phase diagram.
Usually, a disordered initial condition is unable 
to probe the most compact phases. 
On the other hand, high-density compact configurations of 
particles of arbitrary shape are not readily envisageable and 
unexpected features may arise.
In this respect, hard (sphero-)cylinders seem an exception;\cite{Bolhuis97}
hard ellipsoids, thought  for long to crystallise in 
a "stretched-fcc" structure,\cite{Frenkel85} 
were recently shown to also do otherwise.\cite{Donev04}
In order to cope with this problem, 
we have exploited an isopointal search method \cite{Hudson08} 
to construct compact configurations that 
we then used as initial configurations in most of the simulations. 
The method hinges on a structural search for dense packing, 
supported and guided by crystallographic inputs helping 
to reduce its computational cost, 
and coupled to an annealing scheme that progressively increases 
the density of 
a small number of helices within a unit cell, 
until the maximum possible packing is achieved.

\begin{figure}[htbp]
\begin{center}
\includegraphics[width=5.0cm]{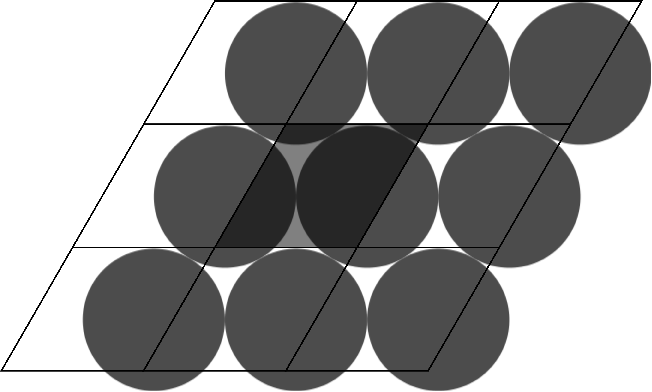} 
\hfill
\includegraphics[width=5.0cm]{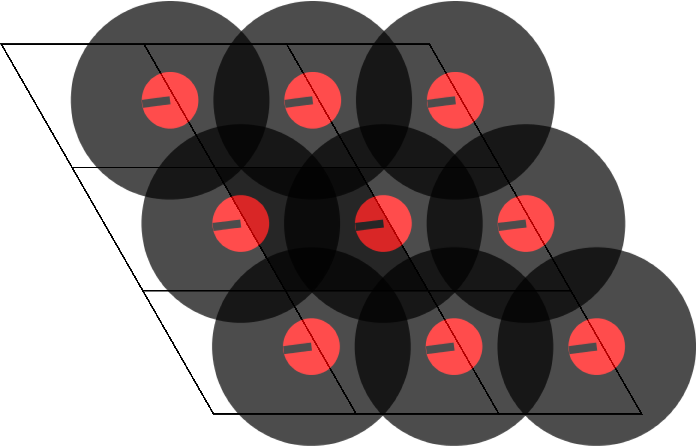} 
\end{center}  
\caption{Maximally packed structure for two different cases.
(Top) Non-overlapping circles, the two-dimensional counterpart of 
hard cylinders (limit case of helices with $r=0$ and $p=\infty$,  
having $\eta_{hex}=\pi/\sqrt{12}$).
(Bottom) 
Overlapping circles corresponding to hard helices with $r=0.2$ and $p=3$.
}
\label{fig:fig2}
\end{figure}

Calculations were made for a single layer of parallel helices with 
their centres of mass lying on the same plane.
This led to 
a considerable simplification in that 
one could limit the analysis to 
the $17$ two-dimensional wallpaper space groups, 
rather than having to deal with 
the full set of the $230$ three-dimensional space groups.
When applied to a system of hard helices with 
radii ranging in the interval $0.1 \le r \le 1$ and 
pitches in the range $1\le p \le 10$,  
the ISM predicts that, 
apart from the peculiar case of $p=1$, 
a specific wallpaper group  with a single helix per unit cell 
provides the maximum possible packing fraction.
Fig. \ref{fig:fig2} (bottom panel) provides a top view of 
the resulting structure in the case $r=0.2$ and $p=3$, 
with the circles having their centre 
coincident with the projection on a perpendicular plane to the helix axis, 
and their radius equal to the helix radius $r$ (the red circle) and 
to $r + 1/2$ (the black circle). 
The black thick lines inside the red circles give the orientation of the  
twofold (C$_2$) symmetry axis of the helices. 
Notice that they are aligned along the same direction,  
meaning that helices are all in register. 
The darker grey areas indicate overlapping regions, 
where the grooves of neighbouring helices intrude into each other voids. 
For any given helix morphology, 
the method provides the shape and area of the unit cell, 
also displayed in Fig. \ref{fig:fig2}.   
The crucial quantity provided by the calculation is 
the ratio $A_{helix}/A_{cell}$
where $A_{helix}$ is the area occupied by the helix 
(i.e. the section of the cylinder containing the helix)
and $A_{cell}$ the area of the unit cell. 
Because of the significant overlap between 
neighbouring helices, this ratio might exceed unity, 
as clearly indicated by the differences in 
the two pictures in Fig.\ref{fig:fig2}.
In case of no overlap, 
the maximal packing would be the bidimensional hexagonal, 
having a packing fraction
$\eta_{hex}=\pi/\sqrt{12}=0.9064\ldots$ and
with each larger circle inscribed 
in the unit cell of area $A_{cell,hex}=(2r+1)^2 \pi/3$ (Fig.\ref{fig:fig2} top). 
This is significantly different from 
the reported example of a helix with 
$r=0.2$ and $p=3$  (Fig.\ref{fig:fig2} bottom) where 
the black circle covers a surface substantially larger than 
the corresponding unit cell.  
The result for $A_{helix}/A_{cell}$  can then be translated into a 
volume fraction as
\begin{eqnarray}
\label{eq:volume_fraction_3d}
\eta&=& \frac{n_{helix} v_{0}}{A_{cell} \left(\Lambda+D\right)}
\end{eqnarray}
where $n_{helix}$ is the number of helices in the unit cell 
($=1$ in the large majority of the cases, as anticipated), 
$v_{0}$ is the volume of the helix, calculated as in Ref.\cite{Frezza13}, and 
$\Lambda$ is the  Euclidean length measured as the component parallel to the main axis of the helix of 
the distance between the first and the last bead.\cite{Frezza13,note3}

\begin{figure}[htbp]
\begin{center}
   \includegraphics[width=9.0cm]{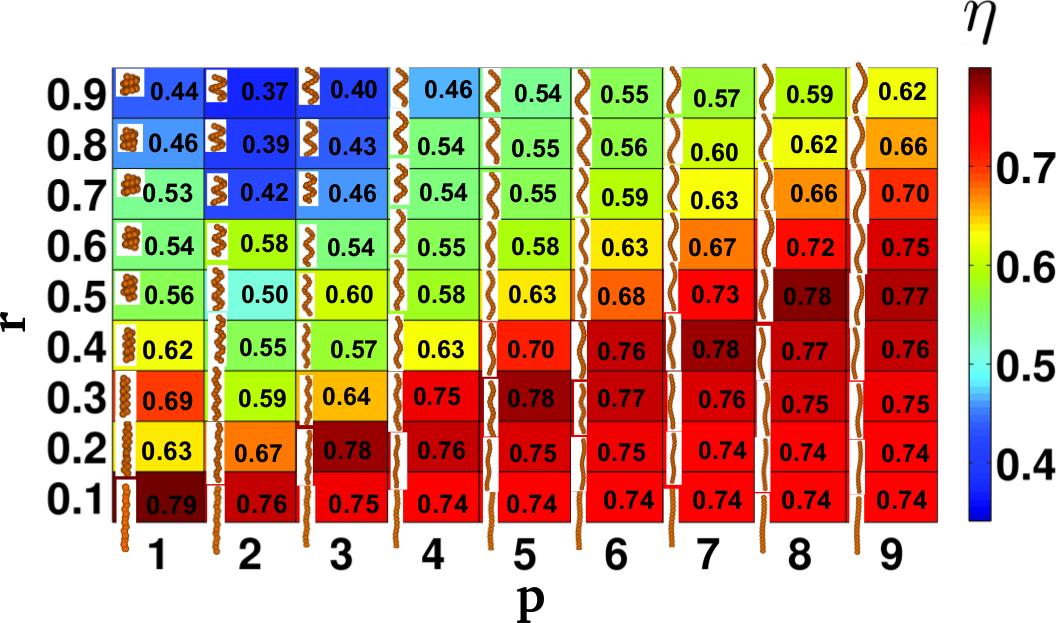} 
\end{center}  
\caption{Colour map of the maximal packing fraction ($\eta$) 
as function of the helix $r$ and $p$. 
The digit inserts indicate the  value of $\eta$. 
A snapshot of the correspoding helix is also shown in the inset.}
\label{fig:fig3}
\end{figure}

The value of $\eta$ is reported in Fig. \ref{fig:fig3}  
in a colour map  as a function of helix $r$ and $p$.
Since these values for $\eta$ have been obtained by considering 
each layer as independent, 
they can only be regarded as a reasonable lower bound of the real maximally packed configuration
that could be achieved by a further interlayer occupancy optimization. 
Yet, the so-built configurations constitute 
a very handy compact initial condition to achieve 
well equilibrated high density structures, as we will see below. 
\subsection{Isobaric Monte Carlo simulations}
The MC-NPT method \cite{Wood68} was used for 
calculating the equation of state of a system of hard helical particles.
Up to $N$=2000 particles were inserted in a
generally triclinic and floppy (i.e. shape adapting) computational box,
with standard periodic boundary conditions. 
Such conditions are fully appropriate as long as 
the spatial periodicity of the mesophases are
comparable with the particle length scale, 
as it is the case of the various phases that 
we will be discussing in the present study. 
They would  \textit{not} be appropriate, however, for 
phases with a periodicity much longer than the particle size. 
This is in general the case of the cholesteric phase,\cite{Priestley74,deGennes93} 
which for this reason cannot be observed in our simulations. 
However, just because of its long length scale, 
the absence of the twist distortion is not expected 
to substantially affect 
the boundaries and the local structure of the nematic phase. 
For this reason, we will henceforth always refer to 
an untwisted conventional nematic phase, 
in spite of the chiral nature of helical particles. 
We will return to this point   later on. 

In the majority of cases, simulations were started from 
a compact configuration as generated by the ISM. 
Few tests were additionally carried out to check 
the robustness of the obtained results with respect to 
the choice of the initial conditions. 
Every simulation run was organised in cycles,
each of them consisting of 
$N/2$ translational and
$N/2$  rotational trial moves, 
performed either using quaternions or the Barker-Watts method \cite{Allen87}
supplemented with a rotation around the helix main axis,
and an attempt to vary shape and volume of the computational box.
The typical length of the equilibration runs was 
$3-4 \times 10^6$ MC cycles.
Equilibration runs were then followed by 
typically $2 \times 10^6$  MC cycle long production runs, 
during which averages of 
several order parameters and correlation functions were accumulated.
These quantities were used to characterise and distinguish 
the various phases.

Dedicated additional simulations were also carried out with 
the helix long axes constrained along a fixed direction 
to validate the theory for the
nematic--to--screw-nematic phase transition described
 in Section \ref{sec:DFT}. 
\subsection{Order parameters and correlation functions}
\label{IIC}
\begin{figure}[htbp]
\begin{center}
   \includegraphics[width=5.0cm]{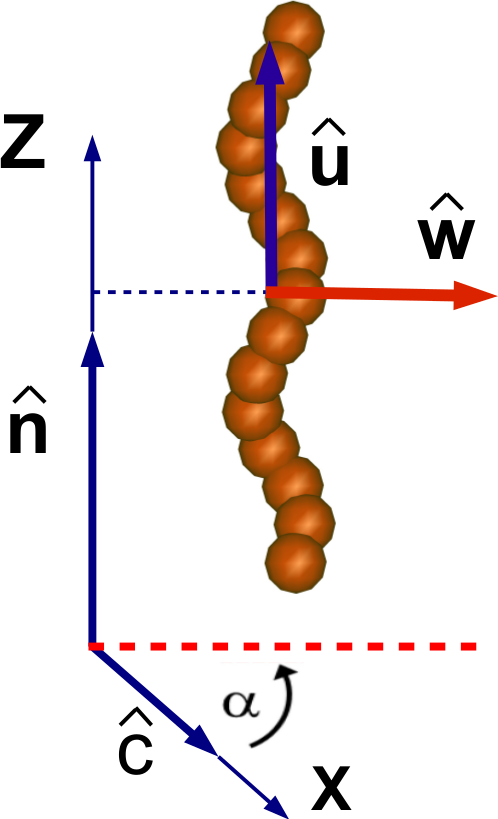} 
\end{center}  
\caption{
Helix with arrows showing the unit vectors $\widehat{\mathbf{u}}$ and  $\widehat{\mathbf{w}}$ 
defined in the molecular frame, and the unit vectors
$(\widehat{\mathbf{n}}, \widehat{\mathbf{c}})$ defined in the laboratory frame.
$X$, $Z$ are the axes of the laboratory frame, with  
$\alpha$ the angle between $\widehat{\mathbf{w}}$ and the $X$ axis. 
}
\label{fig:fig4}
\end{figure}

Different liquid crystal phases will be distinguished by using appropriate
order parameters combined with suitable correlation functions. 
For the definition of these quantities, we refer to  Fig.\ref{fig:fig4} where 
the helix main axis $\widehat{\mathbf{u}}$ and 
secondary axis $\widehat{\mathbf{w}}$, parallel to the C$_2$ symmetry axis,
are shown, along with 
the unit vectors parallel to the main phase director ($\widehat{\mathbf{n}}$) and 
to the minor phase director ($\widehat{\mathbf{c}}$)
at a given position. The various order parameters calculated are as follows. 
\begin{itemize}   
\item  The nematic order parameter \cite{Priestley74,deGennes93} is defined as the averaged second Legendre polynomial
\begin{eqnarray}
\label{sec2:eq2}
\left \langle P_2 \right \rangle &=& 
\left \langle \frac{3}{2} (\widehat{\mathbf{u}} \cdot \widehat{\mathbf{n}})^2 - \frac{1}{2} \right \rangle,
\end{eqnarray}
where the average is over the configurations. 
It can be computed using the standard procedure patterned after 
the early work of Veilliard-Baron \cite{Veilliard-Baron74} by introducing,
for each configuration, the second-rank tensor
\begin{eqnarray}
\label{sec2:eq3}
Q_{\alpha \beta} &=& \frac{1}{N}  \sum_{i=1}^N \frac{3}{2} \widehat{u}_{\alpha}^{i} \widehat{u}_{\beta}^{i} - \frac{1}{2} \delta_{\alpha \beta}
\end{eqnarray}
with $\widehat{u}_{\alpha}$, $\widehat{u}_{\beta}$ the Cartesian components of  $\widehat{\mathbf{u}}$
and $\delta_{\alpha,\beta}$ the Kr\"{o}necker symbol.  
This second-rank traceless tensor is then diagonalized to compute 
the largest eigenvalue and 
the corresponding eigendirection. The latter is
identified as the configuration's nematic director $\widehat{\mathbf{n}}$.
The maximum eigenvalues are then averaged over the configurations,
to give the order parameter $\langle P_2 \rangle$ defined in Eq.(\ref{sec2:eq2}). 
This order parameter (essentially) vanishes in the isotropic phase (I) whereas 
in the nematic phase is (distinctly) larger than zero and 
approaches unity as density increases. 

\item The screw-nematic order parameter \cite{Kolli14,Manna07} is defined as

\begin{eqnarray}
\label{sec2:eq4}
\left \langle P_{1,c} \right \rangle &=& \left \langle \widehat{\mathbf{w}} \cdot \widehat{\mathbf{c}} \right \rangle.
\end{eqnarray}
This order parameter measures 
the average alignment along a common direction ($\widehat{\mathbf{c}}$) of 
the secondary axes ($\widehat{\mathbf{w}}$) of helices having 
their centre of mass  
on the same plane perpendicular to the main director $\widehat{\mathbf{n}}$.
In the screw-like organisation,  the minor phase director $\widehat{\mathbf{c}}$, 
perpendicular to $\widehat{\mathbf{n}}$, rotates around it in a  helical fashion with a pitch $p$.
To determine this order parameter, we have followed the following procedure.
Firstly for each configuration, after having determined the main director  $\widehat{\mathbf{n}}$ 
as explained in the previous item, an untwisting of $-2 \pi Z_i/p$  around $\widehat{\mathbf{n}}$  is enforced on the coordinates of particles,
where $Z_i$ is the coordinate of the center mass of the $i$-th helix along the $Z$ axis parallel to
$\widehat{\mathbf{n}}$.
Then the quantity $\widehat{\mathbf{w}} \cdot \widehat{\mathbf{c}}$ is calculated for each helix and 
finally $\left \langle P_{1,c} \right \rangle$ is obtained by averaging over all helices and configurations. 
The order parameter $\left \langle P_{1,c} \right \rangle$ thus enables us to distinguish between 
the conventional (N) and the unconventional screw-nematic (N$_{S}^{*}$) phases.

\item The smectic order parameter \cite{Priestley74,deGennes93} is defined as 

\begin{eqnarray}
\label{sec2:eq5}
\left \langle 
\tau_1 
\right \rangle 
&=& \left | \left \langle \mathsf{e}^{i 2\pi \frac{\mathbf{R}\cdot\widehat{\mathbf{n}}}{d}}  
\right \rangle \right |,
\end{eqnarray}
with $\mathbf{R}$ the position of a particle centre of mass and $d$ the optimal layer spacing. 
Evaluated following standard prescriptions (e.g. Ref.\cite{Memmer02,Cifelli06}), 
$\langle \tau_1 \rangle$ indicates the onset of a smectic phase, 
where particles tend to organise 
in layers perpendicular to the director $\widehat{\mathbf{n}}$. 

\item The sixfold bond-orientational (hexatic) order parameter and the average number of nearest-neighbours.
The former order parameter is defined as:
\begin{eqnarray}
\label{sec2:eq6}
\left \langle \psi_{6} \right \rangle &=& \left \langle \frac{1}{N} \sum_{i=1}^N \left \vert \frac{1}{n(i)} \sum_{j=1}^{n(i)} e^{6 \mathrm{i} \theta_{ij}}
\right \vert \right \rangle.
\end{eqnarray}
Here $\theta_{ij}$ is the angle 
that the  $i$,$j$ intermolecular  distance vector forms with a pre-fixed axis 
in a plane perpendicular to $\widehat{\mathbf{n}}$,
while $n(i)$ is the number of nearest-neighbours of molecule $i$ within a single layer. 
As $\langle \tau_1 \rangle$ can only signal the onset of a generic smectic phase,
we will then be using $\langle \psi_6 \rangle$ to probe the onset of hexatic order 
(e.g. \cite{Memmer02,Cinacchi08}).
The piece of information stemming from $\langle \psi_{6} \rangle$ can be supported by computing 
the average number $\langle n \rangle$ of nearest-neighbours within each layer,
that tends to $6$ in the hexatic phase. 
This quantity is computed by averaging $n(i)$ over all helices in a plane, and over all possible configurations.
We remark here that the actual value of $n(i)$ is rather sensitive to the definition of nearest-neighbours distance, 
that has always a certain degree of arbitrariness, especially for hard-body particles, 
and here is taken to be $1.1$.
Both $\psi_{6}$  and $n(i)$  display consistent behaviour for different layers, and 
hence the results for a single, arbitraly chosen layer will be shown below.
\end{itemize}

In addition to order parameters, 
we have calculated several positional and orientational correlation functions that 
provide a more detailed picture of a single thermodynamic state point.

\begin{itemize}

\item The parallel positional correlation function:\cite{Cinacchi02}

\begin{eqnarray}
g_{\|} (R_{\|}) &=&
\frac{1} {N} \left \langle \frac{1}{\rho L_x L_y}  
\sum_{i=1}^N \sum_{j \neq i}^N  \delta (R_{\|} - 
\mathbf{R}_{ij} \cdot \hat{\mathbf{n}})  \right \rangle
\label{gpara}
\end{eqnarray}

\item The perpendicular positional correlation function:\cite{Cinacchi02}
\begin{eqnarray}
g_{\perp} (R_{\perp})&=&
\frac{1} {2 \pi R_{\perp} N} \left \langle \frac{1}{\rho L_z}  
\sum_{i=1}^N \sum_{j \neq i}^N  \delta \left(R_{\perp} - 
\left \vert \mathbf{R}_{ij} \times \hat{\mathbf{n}} \right \vert \right)  \right \rangle \nonumber \\
\label{gperp}
\end{eqnarray}

\item The screw-like parallel orientational correlation function:\cite{Kolli14}
\begin{eqnarray}
g_{1,\|}^{\widehat{\mathbf{w}}}(R_{\|}) &=& 
\left \langle \frac{\sum_{i=1}^N \sum_{j\neq i} ^ N \delta(R_{\|}-\mathbf{R}_{ij}\cdot \hat{\mathbf{n}}) (\widehat{\mathbf{w}}_i \cdot \widehat{\mathbf{w}}_j)} 
{\sum_{i=1}^N \sum_{j\neq i} ^ N \delta(R_{\|}-\mathbf{R}_{ij}\cdot \hat{\mathbf{n}})} \right  \rangle. \nonumber \\
\label{gw}
\end{eqnarray}

In Eqs. \eqref{gpara}-\eqref{gw}, $\rho=N/V$ is the number density of the system; 
$V$ is the volume of the sample;
$L_x$ and $L_y$ are the computational box dimensions along
mutually orthogonal directions normal to $\widehat{\mathbf{n}}$; 
$L_z$ is the computational box dimension along $\widehat{\mathbf{n}}$; 
$\delta()$ is the Dirac $\delta$-function and
$\mathbf{R}_{ij}=\mathbf{R}_{j}-\mathbf{R}_{i}$ is the vector joining the centres of helices $i$ and $j$.

The first two positional correlation functions, Eqs. \eqref{gpara} and \eqref{gperp}, are used to 
distinguish a homogeneous (isotropic or nematic) phase (both $g_{\|} (R_{\|})$ and $g_{\perp} (R_{\perp})$ liquid-like), 
a layered (smectic) phase ($g_{\|} (R_{\|})$ solid-like and $g_{\perp} (R_{\perp})$ liquid-like), a columnar phase
($g_{\|} (R_{\|})$ liquid-like and $g_{\perp} (R_{\perp})$ solid-like) or a crystalline phase 
(both $g_{\|} (R_{\|})$  and $g_{\perp} (R_{\perp})$ solid-like).
Eq. \eqref{gw}  is used in connection with 
the screw-nematic order parameter, Eq. \eqref{sec2:eq4}, 
for establishing and quantifying the existence of
a screw-like type of order in the system. 
\end{itemize}
\subsection{Density Functional Theory (DFT)}
\label{sec:DFT}
This section describes the general
density functional theory \cite{Wu07} framework that has been used for 
studying the I--N and the N--N$^{*}_{S}$ phase transitions.

Let us consider a pure system of hard helices
whose mechanical state is described by a set of translational and rotational variables.
The former are collected under the symbol $\mathbf{R}$,
while the latter under the symbol $\Omega$. 
The single--particle density function is then denoted as $\rho(\mathbf{R},\Omega)
=\rho(\mathbf{x})$ and normalised such that $\int d \mathbf{x} \rho(\mathbf{x})=N$.

By retaining only the second and third terms  in the virial expansion,
the excess Helmholtz free energy of such a system is given by:
 \begin{eqnarray}
\beta \mathsf{F}^{ex} &=& \frac{1}{2} \int  
d\mathbf{x} d\mathbf{x}^{\prime} \rho(\mathbf{x}) M( \mathbf{x}, \mathbf{x}^{\prime}) \rho(\mathbf{x}^{\prime})  
 \left [ 1 + \frac{1}{3} \int d\mathbf{x}^{\prime \prime} 
M( \mathbf{x}, \mathbf{x}^{\prime \prime}) \rho(\mathbf{x}^{\prime \prime})  M( \mathbf{x}^{\prime \prime}, \mathbf{x}^{\prime})
\right ],
\end{eqnarray}   
with $\beta=1/k_B T$ and
$M(\mathbf{x}, \mathbf{x}^{\prime})$  
the Mayer function changed in sign.\cite{McQuarrie00}
 
Let us assume that the system can form liquid crystal phases 
and that the translational order, if any, is only present along one direction, $Z$, 
with a periodicity equal to $p$. Thus, the single-particle density function can be expressed as
$\rho(\mathbf{x})=\rho(Z,\Omega)$. 
Therefore the excess free energy density of the system is given by:
\begin{eqnarray}
\frac{\beta \mathsf{F}^{ex}}{V}  = \beta \mathsf{f}^{ex}= 
 \frac{1}{2p}  \int_0^p  d Z d\Omega \rho(Z,\Omega) \int d Z^{\prime} d\Omega^{\prime}  \rho(Z^{\prime},\Omega^{\prime}) 
 \biggl[ a_{excl}( Z, \Omega, Z^{\prime}, \Omega^{\prime}) + 
\frac{1}{3}\int d Z^{\prime \prime} d \Omega^{\prime \prime} \rho(Z^{\prime \prime}, \Omega^{\prime \prime}) 
a_3(Z, \Omega, Z^{\prime}, \Omega^{\prime}, Z^{\prime \prime}, \Omega^{\prime \prime}) \biggr] , \nonumber \\
\label{uno}
\end{eqnarray}
where the functions $a_{excl}( Z, \Omega, Z^{\prime}, \Omega^{\prime})$ and
$ a_3(Z, \Omega, Z^{\prime}, \Omega^{\prime}, Z^{\prime \prime}, \Omega^{\prime \prime}) $ 
have been introduced.
The first is given by:
\begin{equation}
a_{excl}( Z, \Omega, Z^{\prime}, \Omega^{\prime})=
\int \int dX^{\prime} dY^{\prime} M(0,Z,\Omega,X^{\prime},Y^{\prime},Z^{\prime},\Omega^{\prime})
\end{equation}
and is interpreted as the area of the surface obtained by cutting
with a plane perpendicular to the director and at position $Z^{\prime}$
the volume excluded to a particle with orientation $\Omega^{\prime}$ by
a particle at position $Z$ and with orientation $\Omega$.\cite{Cinacchi04} 
The second function in eq.\ref{uno} is given by:
\begin{eqnarray}
&&a_3( Z, \Omega, Z^{\prime}, \Omega^{\prime}, Z^{\prime \prime}, \Omega^{\prime \prime})= \nonumber \\
&&\int  dX^{\prime} dY^{\prime} dX^{\prime \prime} dY^{\prime \prime}
M(0,Z,\Omega,X^{\prime},Y^{\prime},Z^{\prime},\Omega^{\prime}) \nonumber 
M(0,Z,\Omega,X^{\prime \prime},Y^{\prime \prime},Z^{\prime \prime},\Omega^{\prime \prime})
M(X^{\prime},Y^{\prime},Z^{\prime},\Omega^{\prime},X^{\prime \prime},Y^{\prime \prime},Z^{\prime \prime},\Omega^{\prime \prime}),
\end{eqnarray}
but does not lend itself to a ready geometrical interpretation.
Since  $a_{excl}$ and $a_3$  actually depend on the differences 
$\zeta^{\prime}=Z^{\prime}-Z$ and $\zeta^{\prime \prime} = Z^{\prime \prime} - Z$,
the equation above can be re-written in a slightly neater way as:
\begin{eqnarray}
\frac{\beta \mathsf{F}^{ex}}{V}  = \beta \mathsf{f}^{ex}= 
\frac{1}{2p}  \int_0^p  d Z  d\Omega \rho(Z,\Omega) 
\int d \zeta^{\prime} d\Omega^{\prime}  \rho(Z+\zeta^{\prime},\Omega^{\prime}) 
\biggl[a_{excl}( \Omega, \zeta^{\prime}, \Omega^{\prime}) 
+ \frac{1}{3} \int d \zeta^{\prime \prime} d \Omega^{\prime \prime} \rho(Z+\zeta^{\prime \prime}, \Omega^{\prime \prime}) 
a_3(\zeta^{\prime},\Omega,\Omega^{\prime}, \zeta^{\prime \prime}, \Omega^{\prime \prime}) 
\biggr] \nonumber \\
\label{neat}
\end{eqnarray}

The single-particle density function can be decomposed as follows:
\begin{equation}
\rho(Z,\Omega)= \rho(Z) f(\Omega|Z),
\end{equation}
with $\rho(Z)$ the purely translational single particle density, normalised such that 
$(1/p)\int_0^p dZ \rho(Z)=\rho$,
and $f(\Omega|Z)$ the particle orientational distribution function at position $Z$,
normalised such that $\int d\Omega f(\Omega|Z) = 1$, irrespective of $Z$. 
Thus:
\begin{eqnarray}
\frac{\beta \mathsf{F}^{ex}}{V}  &=& \beta \mathsf{f}^{ex}= 
\frac{1}{2p}  \int_0^p  d Z \rho(Z) \int d\Omega f(\Omega|Z) 
\int d \zeta^{\prime} \rho(Z+\zeta^{\prime}) 
\int d\Omega^{\prime} f(\Omega^{\prime}|Z+\zeta^{\prime})  \nonumber \\
&& \biggl[ 
a_{excl}( \Omega, \zeta^{\prime}, \Omega^{\prime}) 
+\frac{1}{3} \int d \zeta^{\prime \prime} \rho(Z+\zeta^{\prime \prime})  f(\Omega^{\prime \prime}|Z+\zeta^{\prime \prime}) 
a_3(\Omega,\zeta^{\prime},\Omega^{\prime}, \zeta^{\prime \prime}, \Omega^{\prime \prime})
\biggr]. 
\label{elibsme}
\end{eqnarray}

In the I, N and N$^{*}_{s}$ phases 
the translational single-particle density does not depend on $Z$, i.e. $\rho(Z)=\rho$. 
Eq. \ref{elibsme} 
can thus be re-written as:
\begin{eqnarray}
\frac{\beta \mathsf{F}^{ex}}{N}  =
\frac{\rho}{2}  \int d \Omega f(\Omega|0)  \int d \zeta^{\prime} \int  d\Omega^{\prime} f(\Omega^{\prime}|\zeta^{\prime}) 
\left [ 
a_{excl}( \Omega, \zeta^{\prime}, \Omega^{\prime}) + \right .
 \left. \frac{\rho}{3}  \int d \zeta^{\prime \prime} \int d \Omega^{\prime \prime}
f(\Omega^{\prime \prime} | \zeta^{\prime \prime}) 
a_3(\Omega, \zeta^{\prime},  \Omega^{\prime}, \zeta^{\prime \prime}, \Omega^{\prime \prime}) \right ].
\label{due}
\end{eqnarray}
If the expansion is truncated at the second virial term, Onsager theory \cite{Onsager49} is recovered. 
One approximate form of the excess free energy density was proposed 
where the expansion  is still truncated at the leading order and 
a pre-factor introduced that is meant to correct for higher order terms:\cite{Parsons79,Lee87} 
\begin{equation}
\frac{\beta \mathsf{F}^{ex}}{N}  =
\frac{G(\eta)}{2} \rho \int d \Omega f(\Omega|0)  \int d \zeta^{\prime} \int  d\Omega^{\prime} f(\Omega^{\prime}|\zeta^{\prime}) 
a_{excl}( \Omega, \zeta^{\prime}, \Omega^{\prime}) 
\label{PL}
\end{equation}
where $G(\eta)=(4 -3\eta)/4(1-\eta)^2$, $\eta$ being the volume fraction, 
equal to $\rho v_{0}$. 
This will be denoted as Parsons-Lee (PL) approximation.
It was originally formulated for monodisperse systems
of hard rod-like particles. Later it was used for
other more complex systems (e.g. Refs. \cite{Cinacchi04,Wensink04}).

The total free energy density contains also 
an ideal term \cite{McQuarrie00} and a contribution accounting for 
the entropy cost of orientational ordering, which is expressed as: 
\begin{eqnarray}
\frac{\beta \mathsf{F}^{or}}{N}  =\int  d \Omega f(\Omega) \ln \left[8 \pi^2 f(\Omega) \right].
\label{Sor} 
\end{eqnarray} 
The orientational distribution function in the I and N phases is independent of $Z$, constant in the former and peaked at $\pm \mathbf{\widehat{n}}$ in the latter. In the N$^*_{s}$ phase it has an implicit dependence on $Z$  because of the local frame rotation around $\mathbf{\widehat{n}}$ with a period equal to $p$. The equilibrium orientational distribution function is 
obtained by functional minimization of the free energy density under the constraint of normalisation.
This leads to the non-linear self-consistent equation: 
\begin{eqnarray}
\ln \left [ K f(\Omega|0) \right] = 
- \rho \int d\zeta^{\prime} \int d\Omega^{\prime} f(\Omega^{\prime}|\zeta^{\prime}) \biggl[a_{excl}( \Omega, \zeta^{\prime}, \Omega^{\prime})
+ \frac{\rho}{2} \int d\zeta^{\prime \prime} \int d\Omega^{\prime \prime}
f(\Omega^{\prime \prime }|\zeta^{\prime \prime}) a_3(\Omega, \zeta^{\prime},  \Omega^{\prime}, \zeta^{\prime \prime}, \Omega^{\prime \prime}) \biggr], 
\label{sc} 
\end{eqnarray} 
with $K$ ensuring that $f(\Omega)$ is correctly normalised.
Once $f(\Omega)$ is known, 
thermodynamic properties such as pressure and chemical potential are obtained by 
differentiating the free energy.
 
We determined the I--N coexistence for helices using Onsager theory with PL correction. Since the orientational distribution function in the I and N phases is independent of the position, $f=f(\Omega)$, eq. \ref{PL} takes the form:\cite{Frezza13} 
\begin{equation}
\frac{\beta \mathsf{F}^{ex}}{N}  =
\frac{G(\eta)}{2} \rho \int d \Omega f(\Omega)  \int  d\Omega^{\prime} f(\Omega^{\prime}) 
v_{excl}( \Omega, \Omega^{\prime}) 
\label{PL2}
\end{equation}
where $v_{excl}( \Omega,  \Omega^{\prime})=\int d \zeta^{\prime} a_{excl}( \Omega, \zeta^{\prime}, \Omega^{\prime})$ is the excluded volume.
A modified form of the Parsons-Lee factor was adopted, as proposed in ref.  \cite{Varga00} for non-spherical particles, using the helix volume $v_0$ calculated as in ref. \cite{Frezza13}.
Numerical minimisation of the Helmholtz free energy was performed
under the constraint of equal pressure, $P$, and chemical potential, $\mu$, in the two coexisting phases:
$P_I = P_N$,  $\mu_I = \mu_N$.
Starting from one point in the isotropic (low density) and one in the nematic phase (high density), calculations at increasing and decreasing density, 
respectively, were performed. 
Coexistence was then identified by the crossing of the curves 
for the I and N branches in the  ($P$, $\mu$) plot.

For the second order N--N$^{*}_{s}$ phase transition we assumed perfect orientational ordering, 
which can be justified by the fact that  in MC simulations this phase transition is observed at very large values of the nematic order parameter. 
In this approximation, the functions $a_{excl}$ and $a_3$ in eq.\ref{due} depend, respectively on 
$(\zeta^{\prime}, \gamma^{\prime})$ and
$(\zeta^{\prime},\gamma^{\prime}, \zeta^{\prime \prime}, \gamma^{\prime \prime}$), with
$\gamma^{\prime}$ ($\gamma^{\prime \prime}$) being the angle between the $\widehat{\mathbf{w}}$ axes of two helices
whose centres of mass are separated by a distance  
$\zeta^{\prime}$ ($\zeta^{\prime\prime}$) along $\widehat{\mathbf{n}}$.  
Thus eq. \ref{due} becomes:
\begin{eqnarray}
\frac{\beta \mathsf{F}^{ex}}{N}  =
\frac{\rho}{2}  \int d \tilde\gamma f(\tilde\gamma|0)  \int d \zeta^{\prime} \int  d\tilde\gamma^{\prime} f(\tilde\gamma^{\prime}|\zeta^{\prime}) 
\left[ a_{excl}( \zeta^{\prime}, \gamma^{\prime}) + 
\frac{\rho}{3}  \int d \zeta^{\prime \prime} \int d \tilde\gamma^{\prime \prime}
f(\tilde\gamma^{\prime \prime} | \zeta^{\prime \prime}) 
a_3(\zeta^{\prime},  \gamma^{\prime}, \zeta^{\prime \prime}, \gamma^{\prime \prime}) \right ], 
\label{ddue}
\end{eqnarray}
where the angles $\tilde\gamma$, $\tilde\gamma^{\prime}$, $\tilde\gamma^{\prime \prime}$ define the orientation of the the $\widehat{\mathbf{w}}$ axes of  helices in the laboratory frame, so $\gamma^{\prime}=\tilde\gamma^{\prime} -\tilde\gamma$, $\gamma^{\prime\prime}=\tilde\gamma^{\prime\prime}-\tilde\gamma$.
In turn, in this approximation Eq. \ref{PL} becomes:
\begin{equation}
\frac{\beta \mathsf{F}^{ex}}{N}  =
G(\eta) \frac{\rho}{2}  \int d \tilde\gamma f(\tilde\gamma|0)  \int d \zeta^{\prime} \int  d\tilde\gamma^{\prime} f(\tilde\gamma^{\prime}|\zeta^{\prime}) 
a_{excl}( \zeta^{\prime}, \gamma^{\prime}).  
\end{equation}
Calculations were performed using Onsager theory with and without PL correction, 
as well as with the virial expansion extended to the third order contribution.
The orientational distribution function at various density values was obtained either by 
numerical solution of the integral equation,  Eq. \eqref{sc}, 
by adapting the method of Ref. \cite{Herzfeld84}, or 
by numerical minimisation of the Helmholtz free energy.\cite{Frezza13}
Compared to the calculations at a second-virial level,
the incorporation of the third-virial terms called for
a significant, but still manageable, increase in the computational cost.
Note that it has been assumed that phase pitch coincides with the helix's.
Exploratory calculations were performed in which this constraint was released.
These calculations confirmed that the equilibrium phase pitch coincides with the helix's,
as was indeed plainly expected and as 
MC  simulations were showing.
\section{Results}
\subsection{Phase diagrams from MC-NPT} 
\begin{figure}
\includegraphics[width=8.0cm]{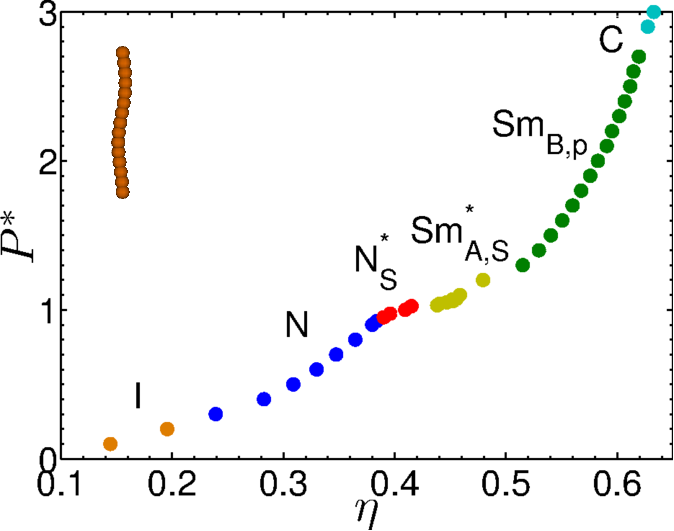}
\caption{
Equation of state for  helices having $r=0.2$ and $p=8$. 
Different colours indicate different phases:  
I=isotropic; N=nematic; N$_{S}^{*}$=screw-nematic; 
Sm$_{A,S}^{*}$=screw-smectic  A; 
Sm$_{B,p}$= smectic B polar. 
}
\label{fig:EOSr0.2p8}
\end{figure}

\begin{figure}
\includegraphics[width=5.0cm]{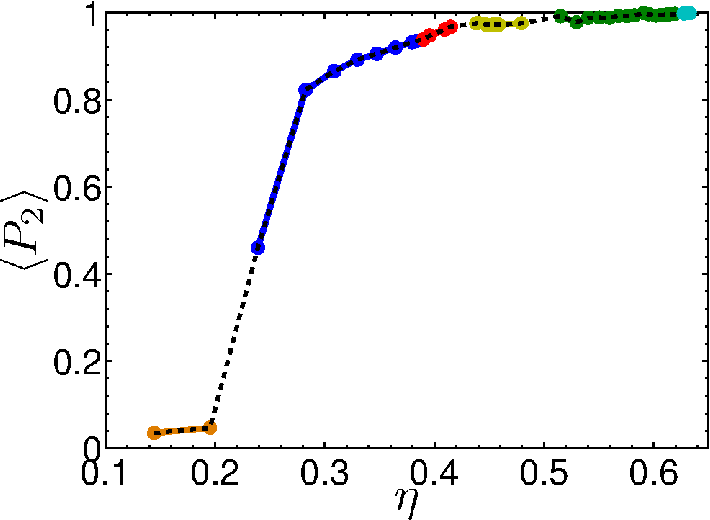} 
\includegraphics[width=5.2cm]{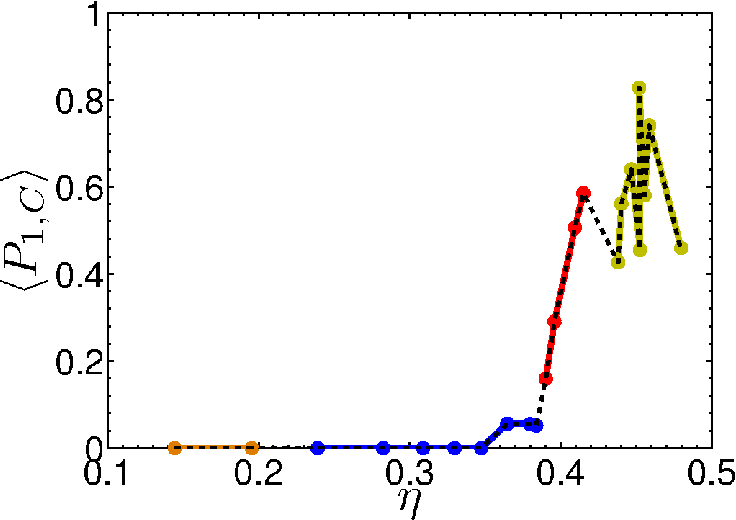}
\caption{
Order parameters $\langle P_2 \rangle$ and $\langle P_{1,c} \rangle$, 
for helices with $r=0.2$ and $p=8$. 
Different colours indicate state points belonging to different phases (see fig. \ref{fig:EOSr0.2p8}).
}
\label{fig:P2_P1c_r0.2_p8}
\end{figure}

\begin{figure}
\includegraphics[width=4.0cm]{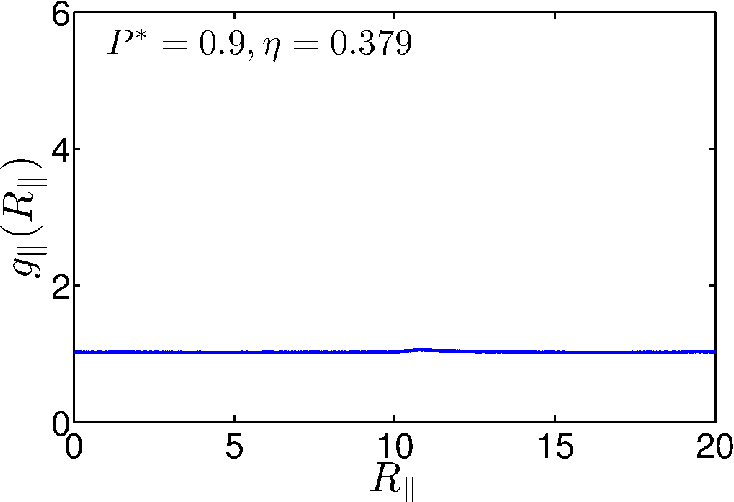} 
\includegraphics[width=4.2cm]{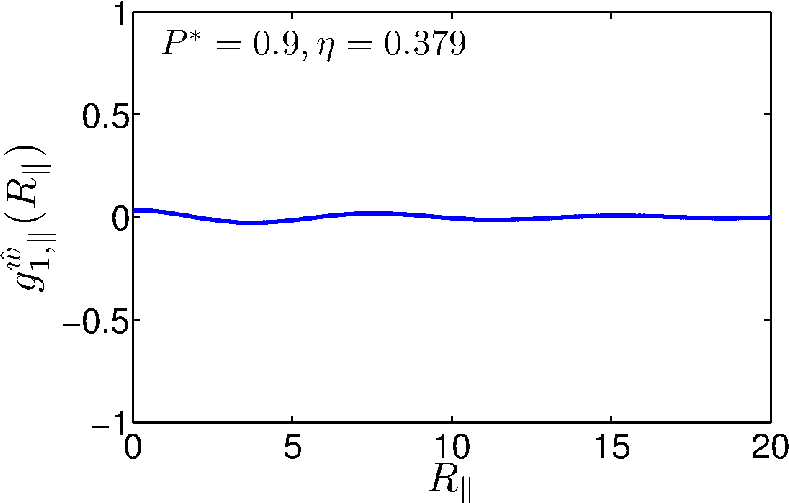}
\includegraphics[width=4.0cm]{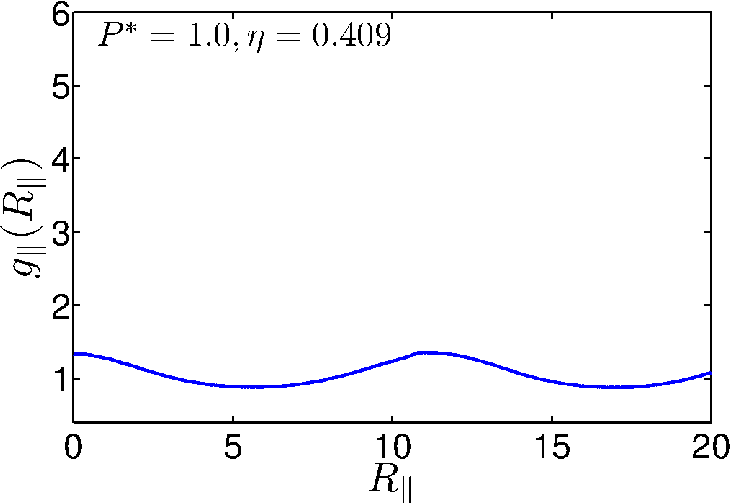} 
\includegraphics[width=4.2cm]{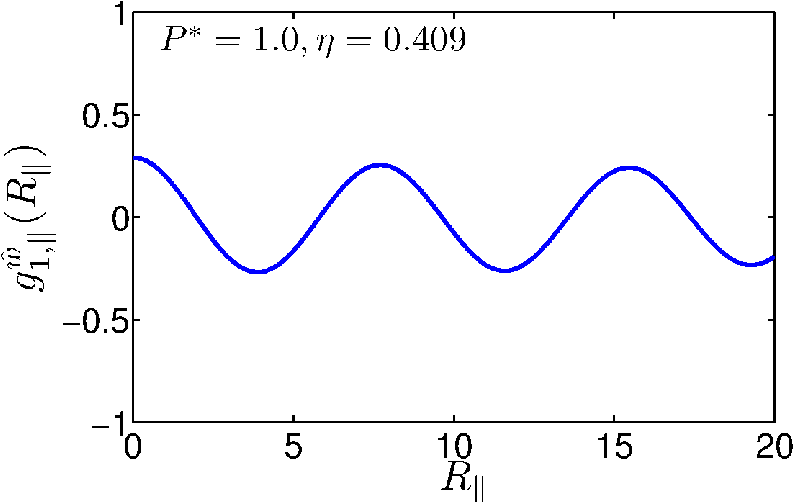}
\includegraphics[width=4.0cm]{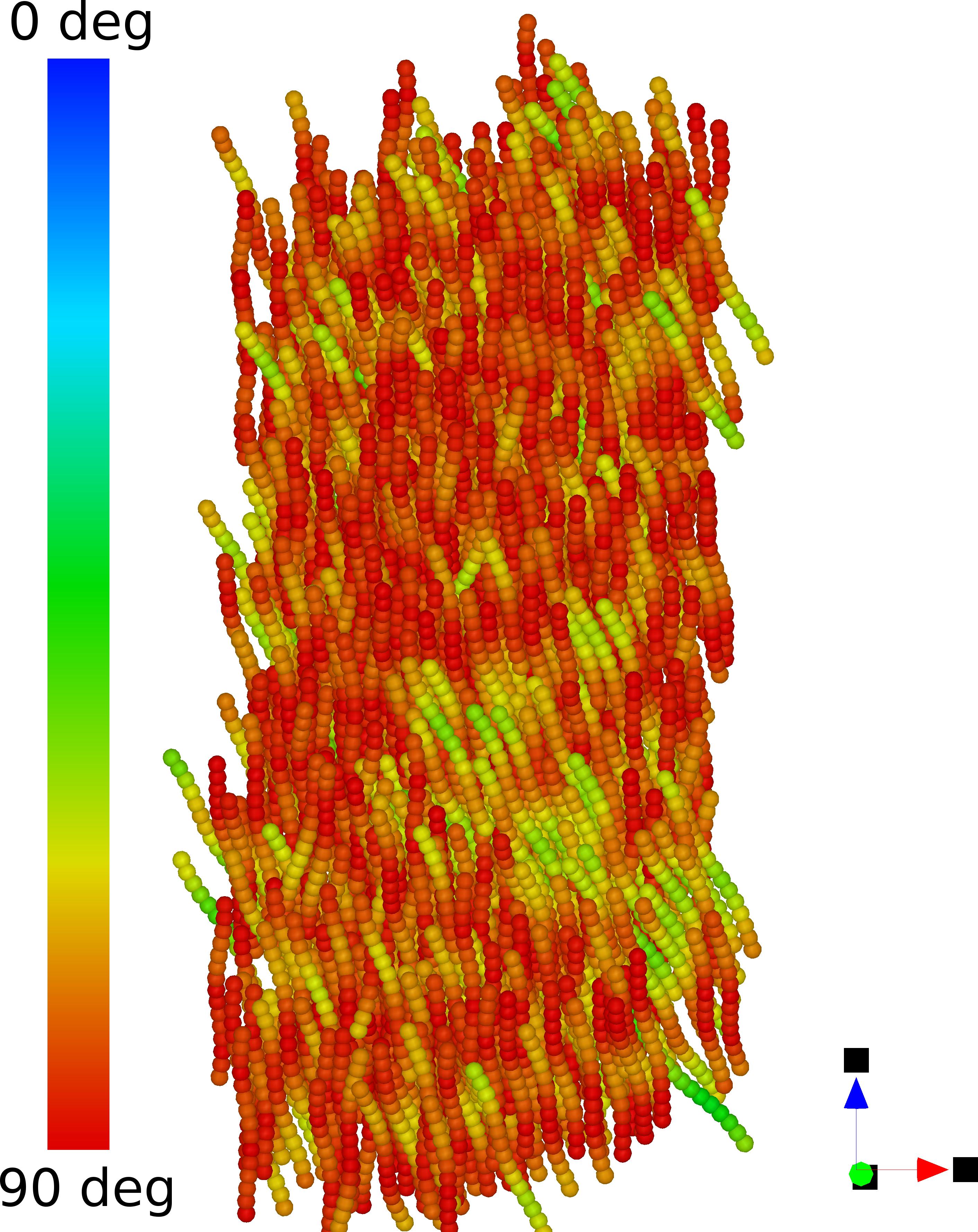} 
\includegraphics[width=4.2cm]{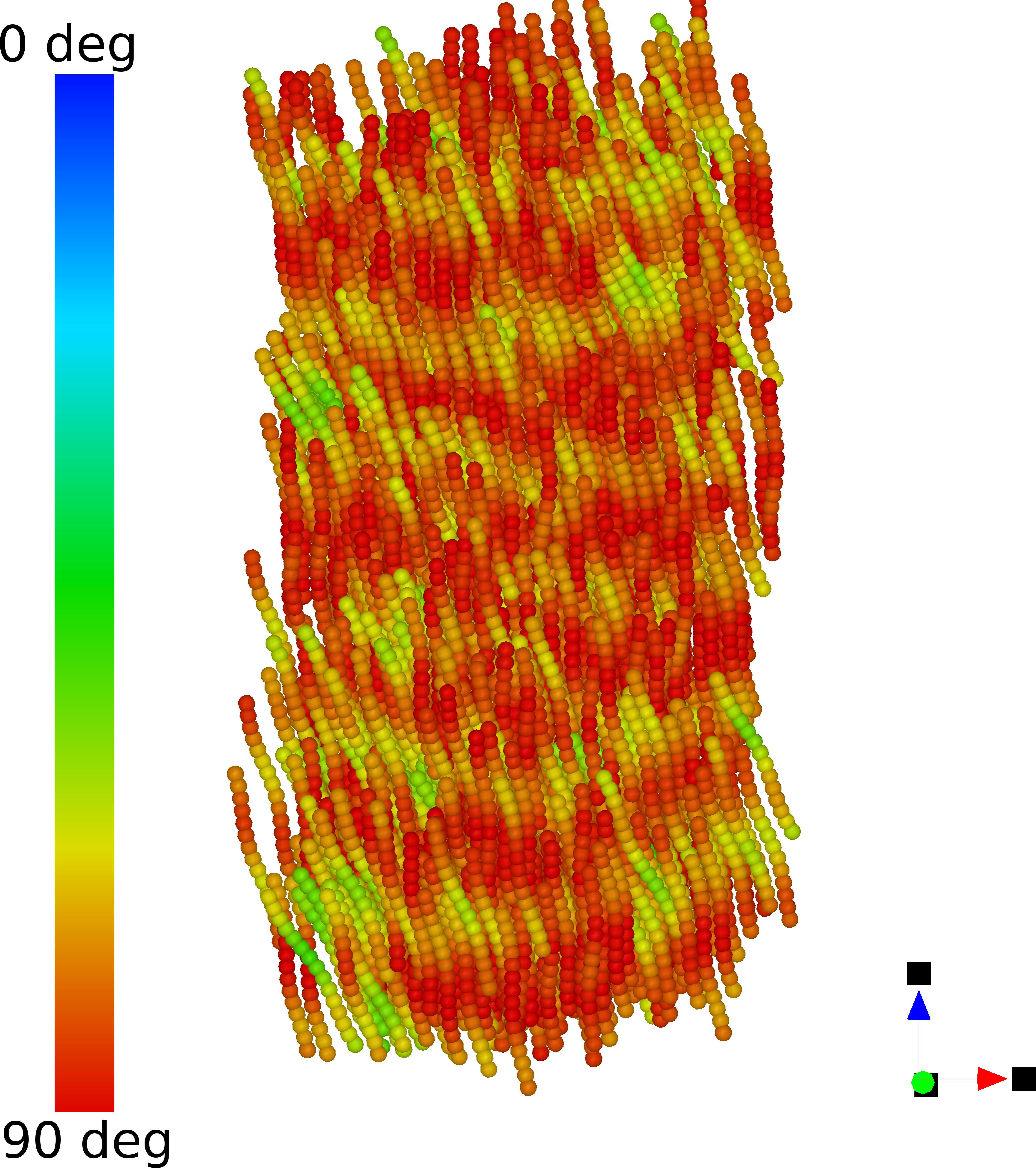} 
\caption{
The $g_{\|} (R_{\|})$ and $g_{1,\|}^{\widehat{\mathbf{w}}}(R_{\|})$ 
correlation functions for helices with $r=0.2$ and $p=8$, 
calculated for $P^{*}=0.9$ in the $N$ phase (left) and 
$P^{*}=1.0$ in the N$_{S}^{*}$ phase (right). 
Also depicted are two corresponding snapshots, 
colour coded according to the local tangent as explained in the text.
}
\label{fig:g_parall_GW1}
\end{figure}

We start by presenting the results obtained for 
the straightest among the helices investigated, 
those corresponding to $r=0.2$ and $p=8$.
Fig. \ref{fig:EOSr0.2p8} shows the equation of state of this system, 
with the reduced pressure $P^* = \beta P D^{-3}$ plotted versus
the volume fraction $\eta=N v_{0}/V$.   
Different phases can be distinguished with the help of 
the order parameters and correlation functions defined
in Section \ref{IIC}.
At low $\eta$ the system is in the I phase, 
but, as $\eta$ approaches a value $\approx 0.23$, 
helices tend to align their long axis ($\widehat{\mathbf{u}}$) along 
a common direction, the main director $\widehat{\mathbf{n}}$.
The onset of the nematic phase is signaled by a jump of 
the order parameter $\langle P_2 \rangle$ to a value $\sim 0.4$, 
as shown in Fig.\ref{fig:P2_P1c_r0.2_p8} (left panel).
This is the conventional N phase, as indicated by 
the absence of translational order and the low or vanishing value of 
all the other order parameters defined in section \ref{IIC}. 
Above $\eta \approx 0.4$, 
Fig.\ref{fig:P2_P1c_r0.2_p8} (right  panel) illustrates how 
the $\langle P_{1,c} \rangle$ order parameter has a marked upswing, 
the signature of screw-like ordering:
the C$_2$ axes of helices ($\widehat{\mathbf{w}}$) tend to 
preferentially align along
a common axis $\widehat{\mathbf{c}}$, 
orthogonal to $\widehat{\mathbf{n}}$ and spiralling around it. 
Unlike the nematic, this order is locally polar, 
i.e. the $\widehat{\mathbf{w}}$ vectors on a  given plane 
perpendicular to $\widehat{\mathbf{n}}$ preferentially point to 
the same direction.
Additional insights on the onset of the  N$_{S}^{*}$ phase are provided by 
the correlation functions 
$g_{\|} (R_{\|})$ and $g_{1,\|}^{\widehat{\mathbf{w}}}(R_{\|})$,
shown in Fig. \ref{fig:g_parall_GW1}, 
for two selected state points corresponding to 
pressure $P^*=0.9$ and $P^*=1.0$, 
across the N--N$_{S}^{*}$ phase transition. 
One can clearly notice the difference in the behaviour of 
$g_{1,\|}^{\widehat{\mathbf{w}}}(R_{\|})$
at the two sides of the phase transition,
with the function at $P^*=1.0$ showing a well developed periodicity that 
matches the pitch of helices $p$.
The sinusoidal  behaviour of $g_{1,\|}^{\widehat{\mathbf{w}}}(R_{\|})$ is 
representative of an azimuthal correlation in planes perpendicular to 
$\widehat{\mathbf{n}}$,
indeed a footprint of the N$_{S}^*$ phase. 
A glance at two snapshots \cite{Gabriel08} also reported in Fig. \ref{fig:g_parall_GW1} gives a visual support of this interpretation.  
Here, as well as in other snapshots reported henceforth, helices are 
colour coded according to their $P_2(\cos\theta)$ value, 
where $P_2$ is the second Legendre polynomial and 
$\theta$ is the angle between the local tangent to helices and 
an arbitrarily chosen axis, not parallel to the main director 
${\widehat{\mathbf{n}}}$. 
This angle changes as the tangent moves along a helix, 
so that $P_2(\cos\theta)$ and then the colour changes, 
with a periodicity equal to half the pitch $p$. 
Therefore, the regular stripes occurring in the bottom right snapshot of 
Fig. \ref{fig:g_parall_GW1} corresponding to 
the  N$_{S}^{*}$ state point ($P^*=1$), 
but absent in the bottom left snapshot corresponding to 
the  N state point ($P^*=0.9$), 
highlight the different organization occurring at the two pressures.

\begin{figure}
\includegraphics[width=4.0cm]{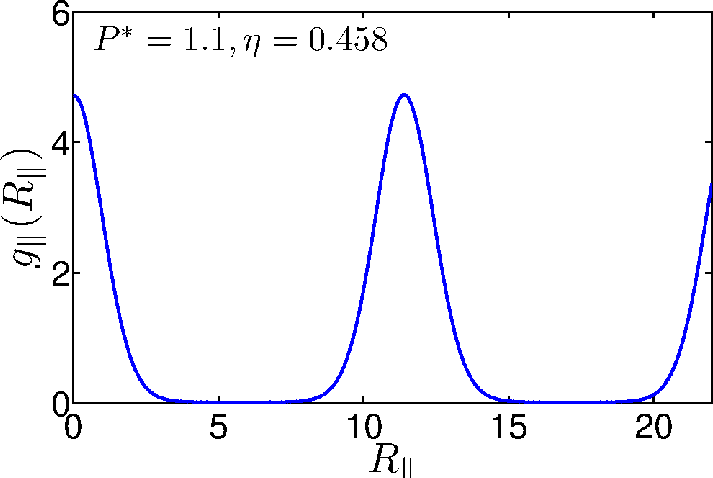} 
\includegraphics[width=4.0cm]{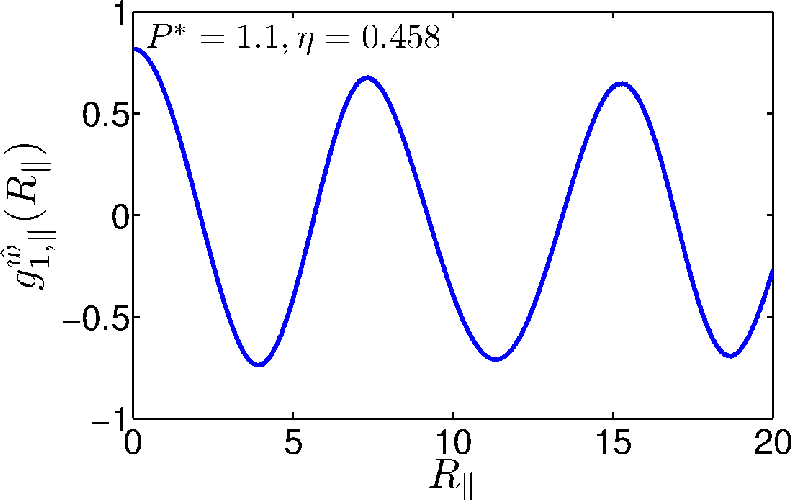}
\includegraphics[width=4.0cm]{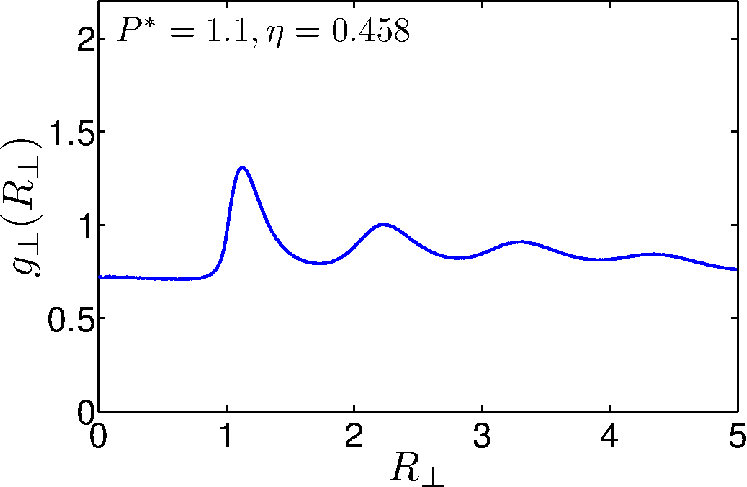}
\includegraphics[width=4.0cm]{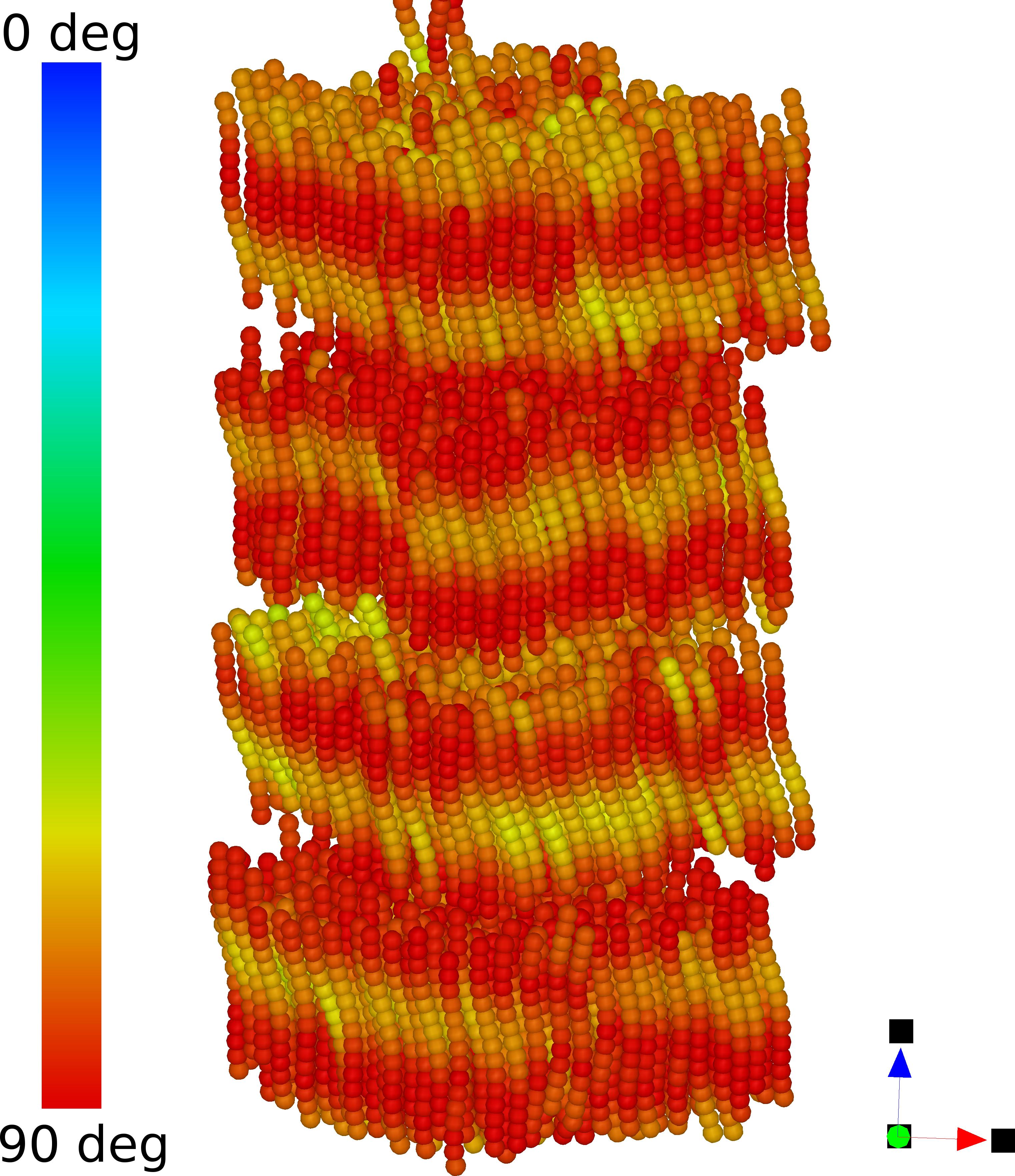} 
\caption{The functions $g_{\|} (R_{\|})$ (top left) and $g_{1,\|}^{\widehat{\mathbf{w}}}(R_{\|})$ (top right) 
at $P^{*}=1.1$, for helices with $r=0.2$ and $p=8$ (Sm$_{A,S}^{*}$ phase..
Also depicted is $g_{\perp} (R_{\perp})$ and 
a corresponding snapshot colour coded according to the local tangent to helices (right). 
}
\label{fig:r0.2p8_P0.9_1.0}
\end{figure}

\begin{figure}
\includegraphics[width=4.0cm]{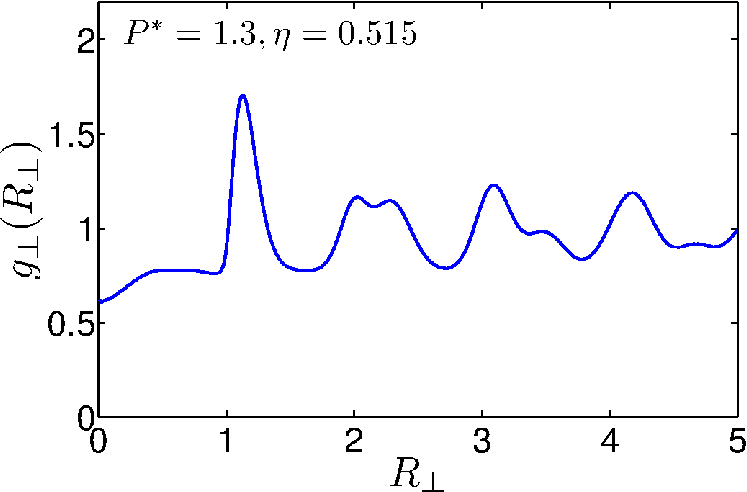} 
\includegraphics[width=4.0cm]{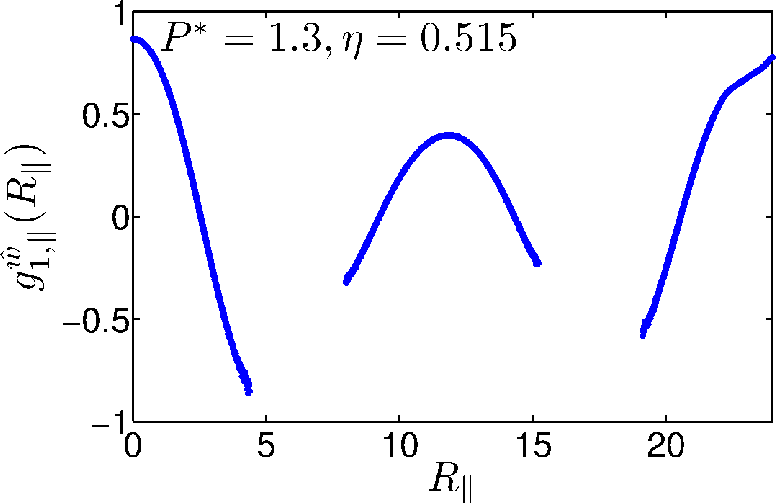} 
\includegraphics[width=4.0cm]{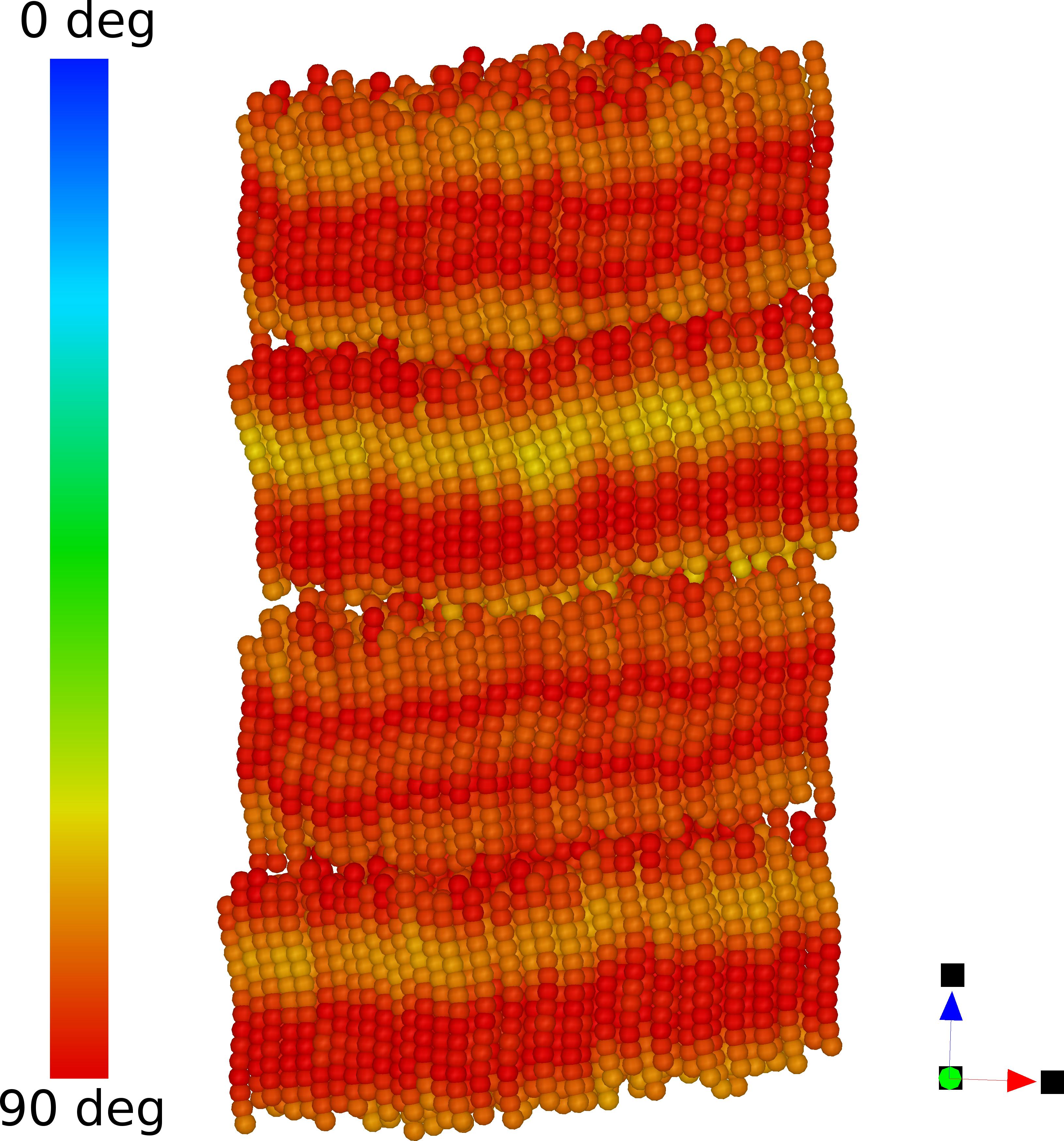} 
\includegraphics[width=4.0cm]{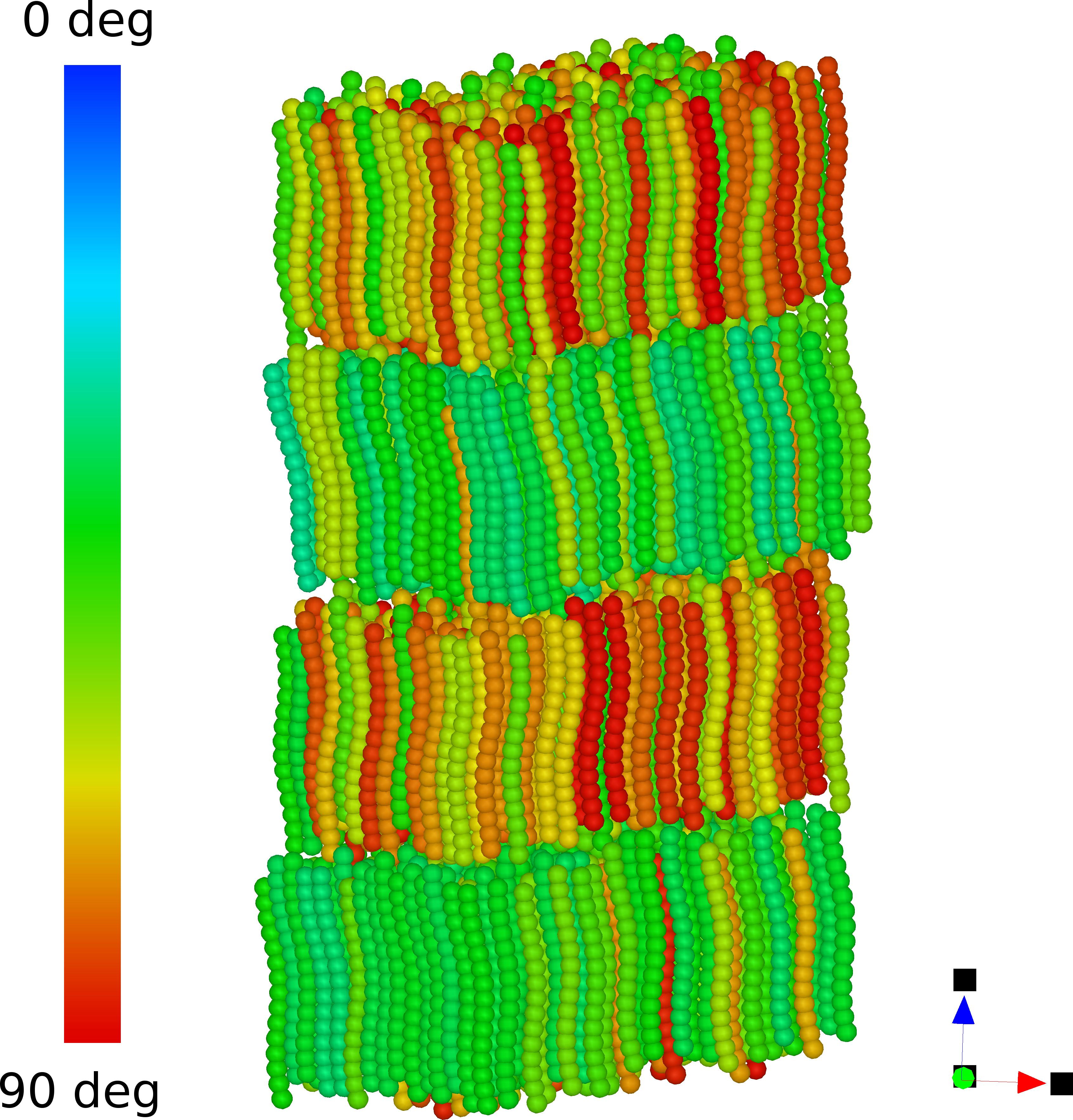}
\caption{Behaviour of $g_{\perp} (R_{\perp})$ (left)  and 
$g_{\|} (R_{\|})$ (right) and 
relative snapshot, 
colour coded according to the local tangent (left) and  
to the C$_2$ axis (right) at $P^{*}=1.3$, 
in the case $r=0.2$ and $p=8$. 
This state point belongs to the Sm$_{B,p}$ phase. 
Note that in the case of the  C$_2$ axis, 
azimuthal rotations of angles in the range $[90^{\circ},180^{\circ}]$ are 
colour coded in the same way as in the case 
$[0^{\circ},90^{\circ}]$ because of the limitations in the QMGA software \cite{Gabriel08}.
For this reason, this color coding will not be further exploited in the remaining of the paper.  
}
\label{fig:r0.2p8_P1.1_1.3perp}
\end{figure}

\begin{figure}
\includegraphics[width=5.0cm]{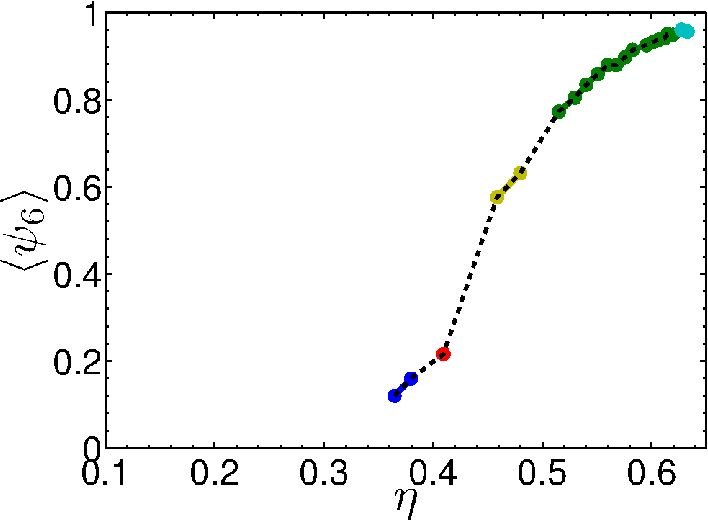} 
\includegraphics[width=5.0cm]{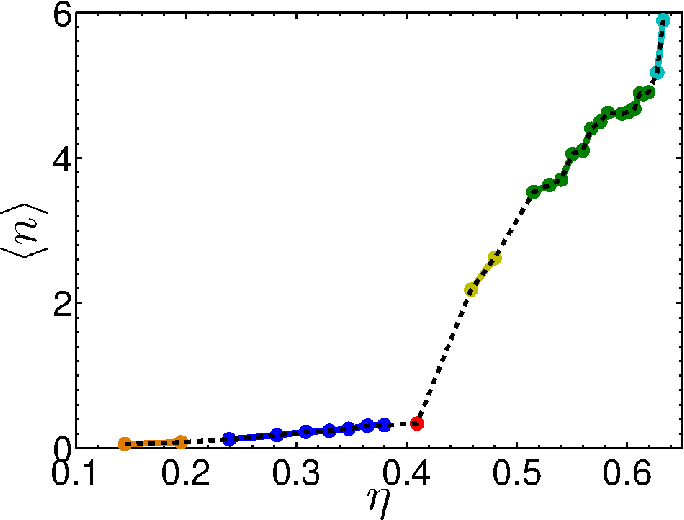}
\caption{
Hexatic order parameter $\langle \psi_{6} \rangle$ and 
average nearest-neigbour number $\langle n \rangle$, 
both indicative of in-plane hexatic ordering, 
for helices with $r=0.2$ and $p=8$. 
State points belonging to different phases are colour   as 
in Fig. \ref{fig:EOSr0.2p8}.
}
\label{fig:Psi6_n_r0.2_p8}
\end{figure}

In Fig. \ref{fig:g_parall_GW1} one can further notice 
a small amplitude oscillation in function $g_{\|} (R_{\|})$  at $P^*=1$, 
that is
absent in the corresponding lower pressure case $P^*=0.9$.
This is indicative of an incipient smectic order, which 
sets in at the slightly higher pressure $P^*=1.1$, 
as confirmed by the  solid-like behaviour of $g_{\|} (R_{\|})$  shown in Fig.\ref{fig:r0.2p8_P0.9_1.0} (top left). 
Here we can recognize  a periodicity of $\sim 12$, 
only slightly longer than the effective length of the helices, 
which is equal to 10.88.
This is different from the periodicity of  
$g_{1,\|}^{\widehat{\mathbf{w}}}(R_{\|})$ (top right),
which corresponds to the helix pitch $p$, here equal to 8.
Thus the screw-like order has combined with 
layer ordering to give rise to a new chiral smectic phase.
The presence of two different periodicities is evident in 
the snapshot in Fig. \ref{fig:r0.2p8_P0.9_1.0}.
The  correlation function $g_{\perp} (R_{\perp})$  
(Fig. \ref{fig:r0.2p8_P0.9_1.0}, bottom left) 
does not provide indication of translational order within each
single layer, and $\langle \psi_{6} \rangle$ is found to be very small. 
This screw-smectic phase,  
globally uniaxial with the main director perpendicular to the layers, 
is of type $A$ and labelled as Sm$_{A,S}^{*}$.
At higher pressure, $P^*=1.3$ ($\eta \approx 0.52$), 
hexatic order sets in within each single layer, as shown by the
behavior of $g_{\perp} (R_{\perp})$ that exhibits well developed characteristic  double peak structure, with maxima at 
$\sqrt 3 \sigma$ and $2 \sigma$ (Fig. \ref{fig:r0.2p8_P1.1_1.3perp}, left panel), $\sigma$
being the position of the main, nearest-neighbour peak.
The presence of hexatic order is further confirmed by the high value of  $\langle \psi_{6} \rangle$ 
(Fig.\ref{fig:Psi6_n_r0.2_p8} left panel) and by the fact that the average number of nearest-neighbours $\langle n \rangle$
tends to $6$ at $\eta \approx 0.52$. 
The plot of  $g_{1,\|}^{\widehat{\mathbf{w}}}(R_{\|})$ (Fig.\ref{fig:r0.2p8_P1.1_1.3perp}, top right panel) shows  a clear in-plane azimuthal correlation, but the absence of the helical periodicity that was present in the  Sm$_{A,S}^{*}$ phase. The correlation now is only within layers, which are uncorrelated from each other. This difference from the Sm$_{A,S}^{*}$   phase clearly appears from comparison of the relative snapshots of Fig. \ref{fig:r0.2p8_P0.9_1.0} with those of Fig.  \ref{fig:r0.2p8_P1.1_1.3perp}.  
We refer to this phase as smectic B polar (Sm$_{B,p}$), to highlight the presence of hexatic order combined with polarity  within layers.
Note that the gaps appearing in the $g_{1,\|}^{\widehat{\mathbf{w}}}(R_{\|})$ 
are indicative of the absence of particles with those 
particular $R_{\|}$ values
and are specific of  state points at very high density. 

\begin{figure}
\includegraphics[width=4.0cm]{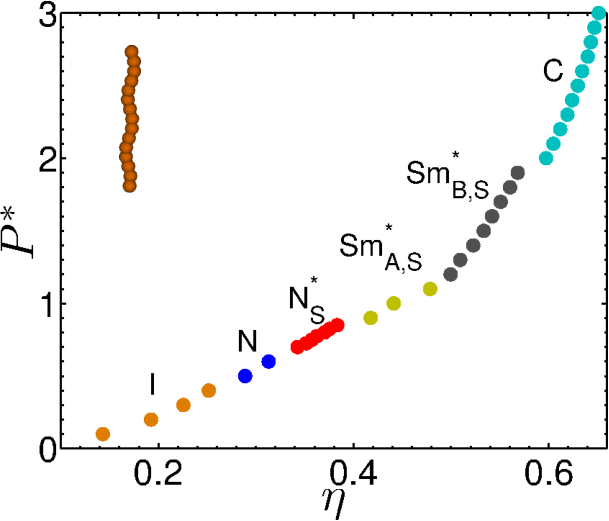} 
\includegraphics[width=4.0cm]{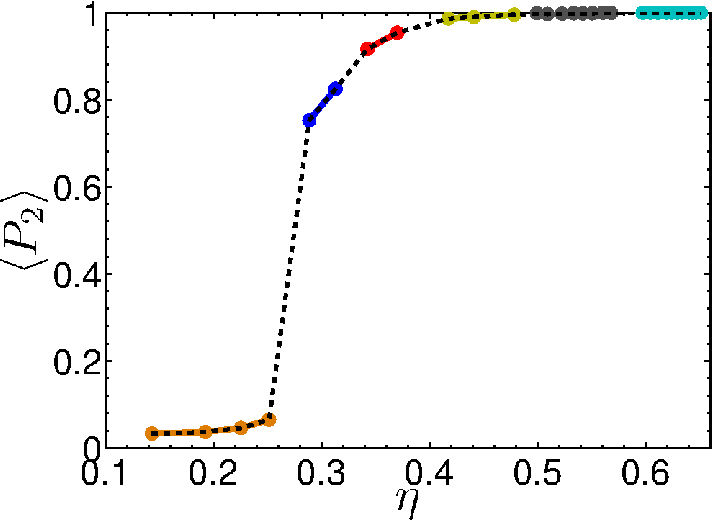} 
\includegraphics[width=4.0cm]{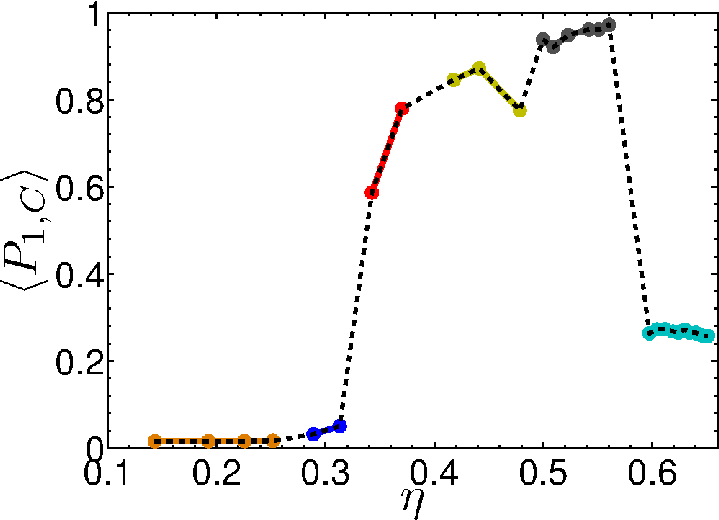} \\
\includegraphics[width=4.0cm]{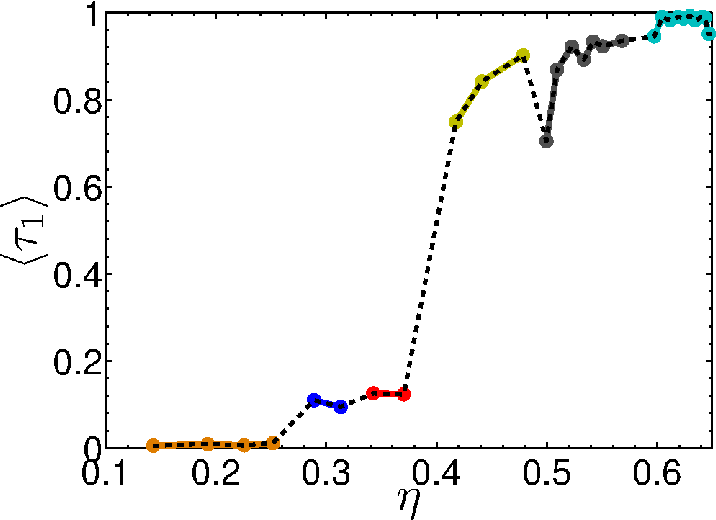} 
\includegraphics[width=4.0cm]{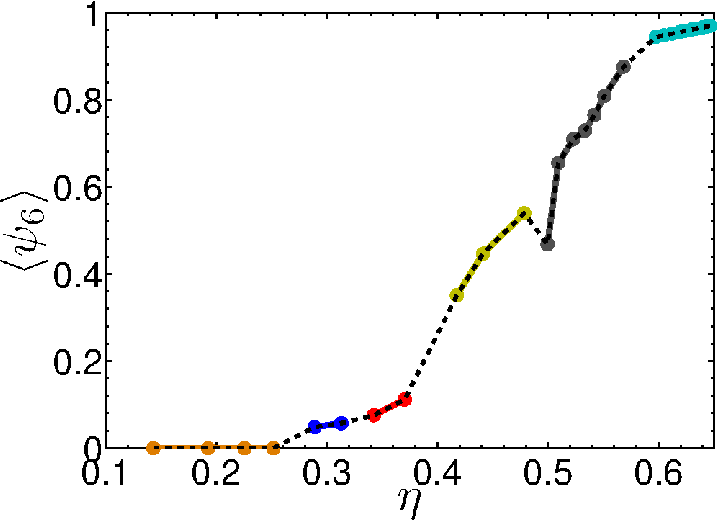}
\includegraphics[width=4.0cm]{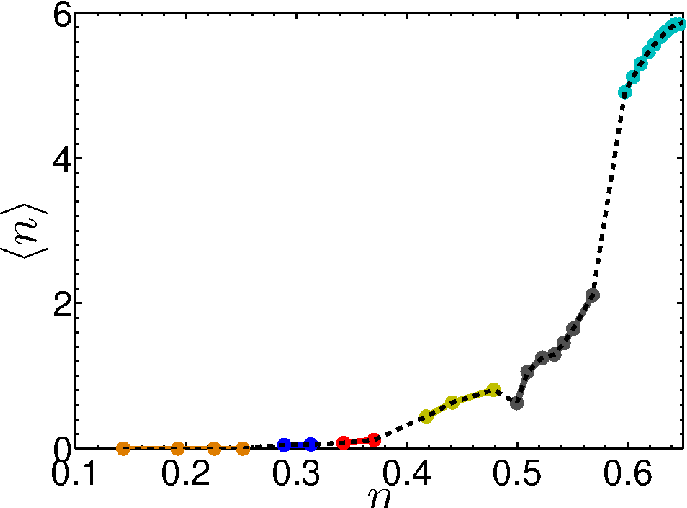} 
\caption{
Equation of state and order parameters $\langle P_2 \rangle$, $\langle P_{1,c} \rangle$,  $\langle \tau_{1} \rangle$, 
$\langle \psi_{6} \rangle$ and average nearest-neigbors $\langle n \rangle$, (in typewriter order) in the case $r=0.2$ and $p=4$.
Note that in the plot for the equation of state (top,left), additional points between the two extrema, have been included in the N$_{s}^{*}$ phase, for completness. 
}
\label{fig:order_parameters_r0.2_p4}
\end{figure}

On increasing the helical twist, the onset of unconventional screw-like phases  becomes more and more pronounced. 
We first keep the radius fixed at $r=0.2$ and decrease the pitch down to $p=4$. 
The equation of state and corresponding order parameters are reported in Fig.\ref{fig:order_parameters_r0.2_p4}. 
As for the case  $r=0.2$ and  $p=8$, there are both the N and the N$_{S}^{*}$ phase,
but the latter becomes predominant in this case. Also the higher density smectic phase exhibits new features, 
with the  $\langle P_{1,c} \rangle$ order parameter being large throughout the entire smectic range and a final sudden drop only at the onset of the compact phase C. 
Thus all smectic phases exhibit screw-like order. 
However they are distinguished by the profile of $g_{\perp} (R_{\perp})$ and by a substantially different behaviour of 
$\langle \psi_{6} \rangle$, which is larger in the state points belonging to the phase denoted as Sm$_{B,S}^{*}$ than the Sm$_{A,S}^{*}$ phase. 
Perhaps surprisingly, the average nearest-neighbour number $\langle n \rangle$ in the Sm$_{B,S}^{*}$ phase remains significantly smaller than $6$, in spite of the
large value of $\langle \psi_{6} \rangle$. As remarked,  this quantity is very sensitive to 
the definition of the nearest-neighbour distance that contains a significant degree of arbitrariness, and hence might be more accurate for some state points than others.
A top view of the relative snapshots none the less confirms the presence of a hexatic ordering in the state points labelled as Sm$_{B,S}^{*}$ and not in those
labelled as  Sm$_{A,S}^{*}$. 

\begin{figure}
\includegraphics[width=8.0cm]{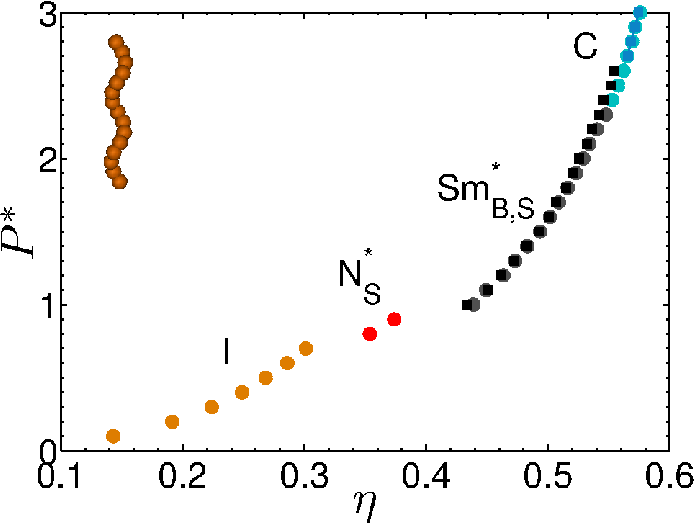} 
\caption{
Equation of state for helices with $r=0.4$ and $p=4$. 
Filled circles 
refer to the same initial conditions used throughout this work, 
whereas filled darker squares refer to 
different initial condition, as detailed in the text.
}
\label{fig:eos_r0.2_p4}
\end{figure}

\begin{figure} 
\begin{center}
\includegraphics[width=6.0cm]{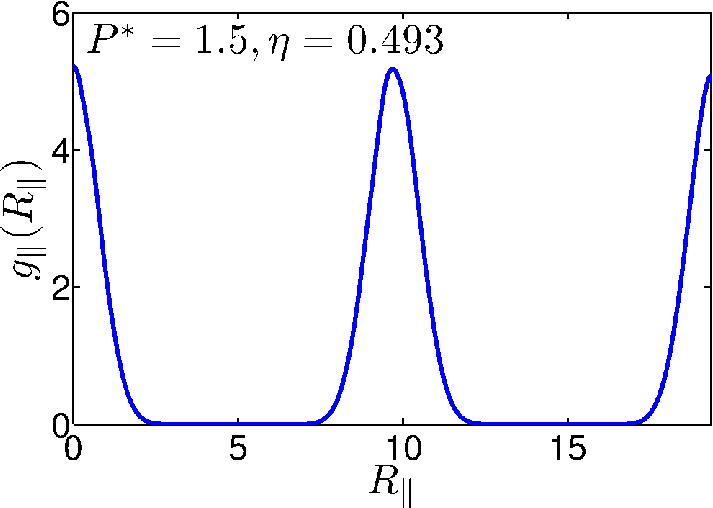} 
\includegraphics[width=6.0cm]{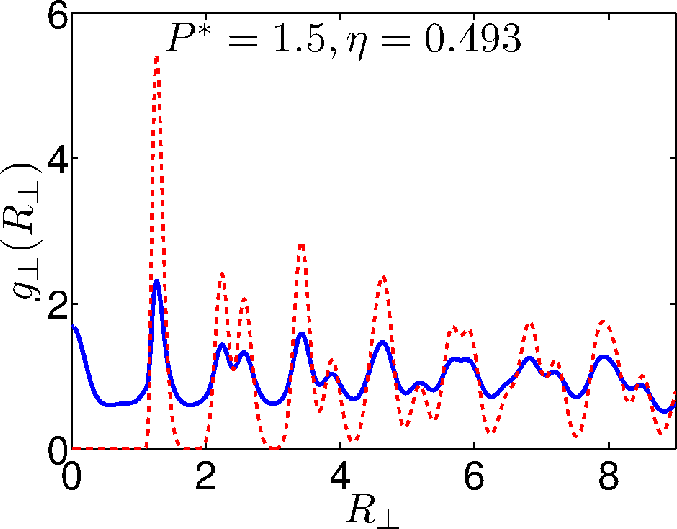} 
\includegraphics[width=6.3cm]{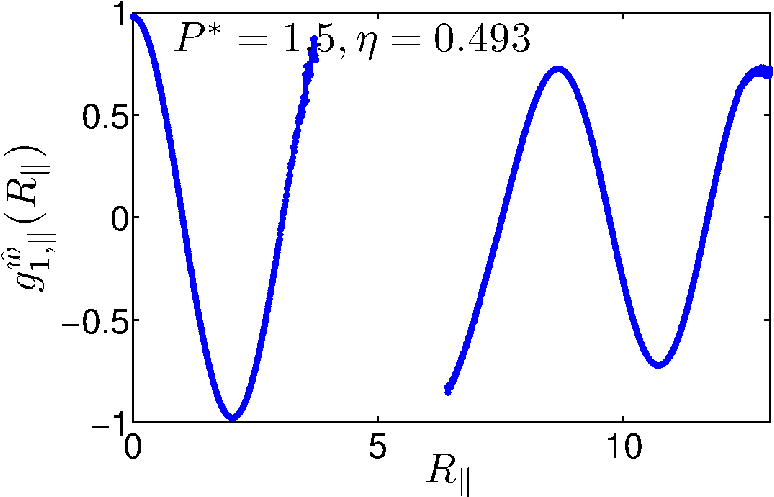} 
\end{center}
\caption{Top left panel. Plot of $g_{\|} (R_{\|})$ in the case of $r=0.4$, $=4$, at reduced pressure $P^{*}=1.5$ (Sm$^\ast_{B,S}$ phase). Top right panel. Profile of $g_{\perp} (R_{\perp})$
calculated by averaging over all layers (solid line) and over a single layer (red dotted line). Bottom panel. Behavior of $g_{1,\|}^{\widehat{\mathbf{w}}}(R_{\|})$. 
}
\label{fig:nuovafigura}
\end{figure}

\begin{figure}
\includegraphics[width=5.0cm]{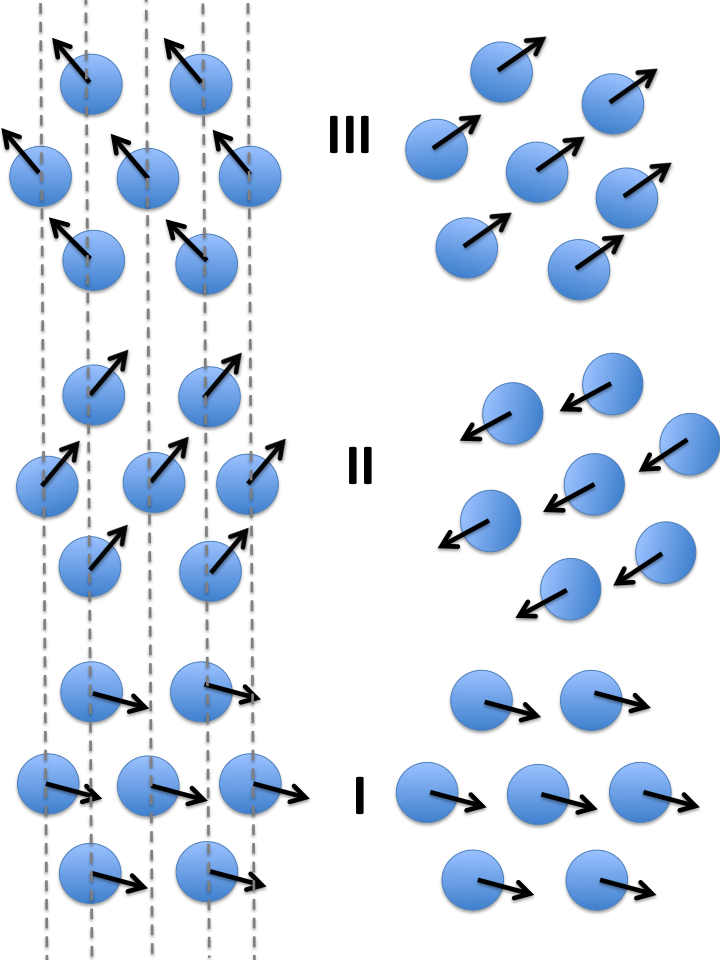} 
\caption{Cartoon of the smectic B phases discussed in the text. Circles represent transversal sections through the center of helices and arrows  the corresponding $\widehat{\mathbf{w}}$ vectors. I,II, III indicate adjacent layers.
Left:  Sm$_{B,S}^{*}$ phase,  with through-layers positional correlation (AAA structure, highlighted by the dashed lines) and screw-like (orientational) correlation of $\widehat{\mathbf{w}}$ vectors. 
Right: Sm$_{B,p}$ phase, with in-layer correlation of $\widehat{\mathbf{w}}$ vectors and neither positional nor orientational correlation between layers.
}
\label{fig:stratiA_vs_stratiB}
\end{figure}

The "curliest"  among the investigated helices are those with  
$r=0.4$ and $p=4$, 
whose morphology is displayed in the inset of Figure \ref{fig:eos_r0.2_p4}
showing the equation of state.
This exhibits important differences with the phase diagram reported in 
Figs. \ref{fig:EOSr0.2p8} and \ref{fig:order_parameters_r0.2_p4},
with only two intermediate high-density liquid crystal phases being present, 
the  N$_{S}^{*}$ and  Sm$_{B,S}^{*}$.  
The system then undergoes a direct first-order transition from  
the  I to the N$_{S}^{*}$ phase,
without an intermediate N phase.
This behaviour can be ascribed to the combined effect of significant twist and small effective aspect ratio,
in agreement with the interpretation given above of the results obtained for the less curly helices.
One more novel feature is a direct transition from the N$_{S}^{*}$ phase to 
a smectic phase with in-plane ordering Sm$_{B,S}^{*}$.
The profile of $g_{\|} (R_{\|})$  obtained at $P^{*}=1.5$ 
(Fig. \ref{fig:nuovafigura} top panel) indicates layering with a periodicity close to the effective helix length, 9.47 in the present case. Hexatic in-plane order is inferred from  the behavior of $g_{\perp} (R_{\perp})$ (Fig. \ref{fig:nuovafigura} central panel) and from the corresponding high value of $\langle \psi_{6} \rangle$. However, differently from  Sm$_{B,p}$ phase  (Fig. \ref{fig:r0.2p8_P1.1_1.3perp}), here  there is additional screw-like ordering, with a period equal to the helix pitch $p=4$, which is evidenced by the correlation function $g_{1,\|}^{\widehat{\mathbf{w}}}(R_{\|})$  (Fig. \ref{fig:nuovafigura} bottom panel) and by the  high $\langle P_{1,c} \rangle$ order parameter. Another interesting feature supporting the Sm$_{B,S}^{*}$ nature of the smectic phase is included in the (red) dotted line of $g_{\perp} (R_{\perp})$ (Fig. \ref{fig:nuovafigura} central panel) that reports the behaviour of $g_{\perp} (R_{\perp})$ 
when 
the average is limited to a single layer. In particular, the absence of the first peak at 
$R_{\perp} \approx 0$ in this case, and conversely present when the average is carried out over all layers, is a clear indication of a AAA structure,
reminiscent of a columnar structure,
where a helix belonging to a given layer locks with the one stacked immediately on its top, and belonging to the successive layer, to form an essentially "infinite" helix spanning the full computational box.
This essential difference between the structure of the Sm$_{B,S}^{*}$ and the Sm$_{B,p}^{*}$ phases is summarized in the sketch of Fig. \ref{fig:stratiA_vs_stratiB}.
While in the Sm$_{B,S}^{*}$ phase (Fig. \ref{fig:stratiA_vs_stratiB} left) helices are azimuthally correlated within each plane and screw-like correlated
between different planes, in the Sm$_{B,p}$ phase (Fig. \ref{fig:stratiA_vs_stratiB} right) only intra-plane azimuthal correlation is present, with
different layers being uncorrelated both positionally and orientationally.
Given the crucial role that starting configurations may have at 
high density,  
as a final point it is instructive to dwell on  
their effect on the final phase diagram. 
We remind at this stage that all results discussed so far were obtained by 
starting from a very compact initial configuration,
obtained by the ISM
and then equilibrated at the appropriate value of $P^*$.
While for the isotropic and nematic phases 
a different initial condition would result into 
an almost indistinguishable picture, 
this is not necessarily true for higher density phases, 
as it also happens in the case of hard spherocylinders.\cite{Bolhuis97} 
This turns out to be also the case here, 
as reported in Fig.\ref{fig:eos_r0.2_p4}, 
where the original results (filled circles) are contrasted with those obtained 
starting from an equilibrated configuration at
the immediately lower pressure (filled squares). 
In both cases the first smectic point  was 
obtained from the original compact configuration.
This small reagion of hysteresis indicates a maximal range of uncertainty of the true thermodynamic coexistence pressure. The results collected so far point to the existence of a first-order transition between the Compact and the Smectic phases: the hysteresis observed may be interpreted as a signal of it.
\subsection{Locating the isotropic-to-nematic phase transition}
It proves of interest, 
at this stage, to discuss how to properly
locate the  volume fractions and pressure at the isotropic--nematic coexistence.
The most direct method 
is a technique   known as Successive Umbrella Sampling (SUS),\cite{Virnau04} 
originally developed for 
the calculation of the gas-liquid coexistence in 
the grand-canonical ensemble. 
In the isotropic-nematic coexistence of hard rods this has been 
discussed in Ref.\cite{Vink05}.
Although it could be clearly applied to the present case as well, we have found this to be particularly
problematic as a result of the combination of the sole hard-core interactions and of the reduced aspect ratio.
As the aspect ratio decreases, the I--N transition shifts to higher densities, and insertion of
a particle becomes increasingly
harder.
This agrees with a similar observation made by the authors of Ref.\cite{Vink05}, who estimated 
$15$ as the minimum aspect ratio to study the phase transition with a reasonable computational effort, 
whereas
our helices have typical aspect ratios of the order of $10$ or less. This notwithstanding, SUS can still be applied
to a helical particle system by using a somewhat more elaborate procedure that will be discussed elsewhere.

\begin{figure}
\includegraphics[width=6.0cm]{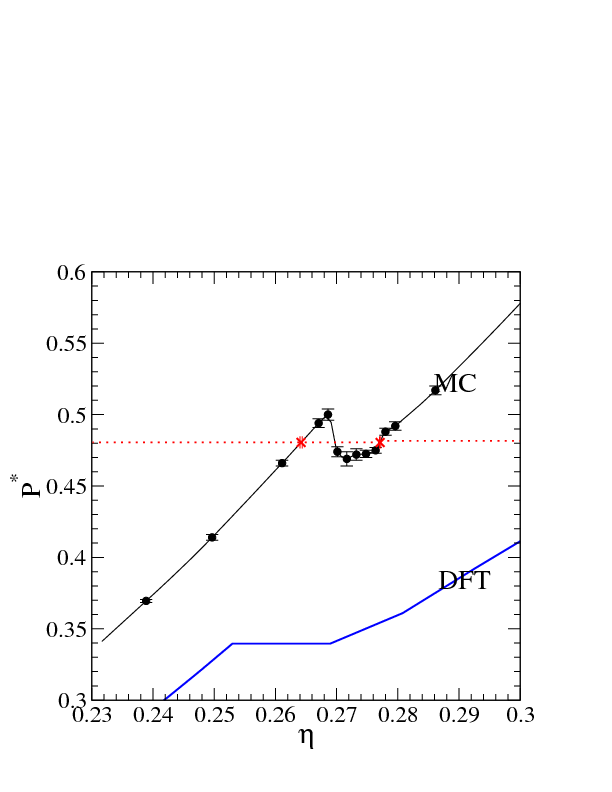}
\caption{
Equal area construction in the case of $r=0.2$ and $p=4$. 
Points are results from MC simulations, traversed by their Akima spline interpolation. 
Resulting values for coexisting volume fractions are 
$\eta_{I}=0.2642 \pm 0.0002$ and $\eta_{N}=0.2772 \pm 0.0001$ (crossed points) at 
pressure $P_{IN}=0.4805 \pm 0.0035$ (dotted line).
Also displayed are the results from DFT (think solid line).
}
\label{fig:maxwell}
\end{figure}

Under these conditions, we have here found it more convenient to resort to a different procedure that, 
albeit less direct,
is still able to provide a rather accurate value of the coexisting densities and pressure.
The basic idea is to perform, via MC-NVT simulations, 
a detailed description of the equation of state 
across the I--N transition, and
then use an
equal area 
construction to infer the coexisting volume fractions and pressure. 
This is depicted  in Fig. \ref{fig:maxwell} in the case $r=0.2$ and $p=4$, 
which is a blow-up of the case analyzed 
in Fig.\ref{fig:order_parameters_r0.2_p4} close to the I--N phase transition.
The Mayer-Wood loop \cite{Mayer65} is consistent with a first order transition, 
and an equal area 
construction provides the two coexisting volume fractions 
$\eta_{I}=0.2642 \pm 0.0002$ and $\eta_{N}=0.2772 \pm 0.0001$ (crossed points) 
at pressure $P_{IN}=0.4805 \pm 0.0035$ (dotted line).
Notwithstanding the finite size effects, it is worth noticing that the precision and reliability 
of this result does not unfavourably compare with those usually obtained via SUS calculations.

These findings can be contrasted with those obtained via the DFT theory, eq. \ref{PL2}, as illustrated in Section \ref{sec:DFT}. 
This result is also reported in Fig.\ref{fig:maxwell}
as a thick solid line. 
Roughly speaking, we find DFT to underestimate the coexistence pressure
by $\approx 15\%$ and the coexistence densities by $\approx 4\%$. This is consistent with previous comparison with NPT simulations \cite{Frezza13}
and with the typical accuracy achieved by DFT calculations.
\subsection{Theoretical description of the nematic--screw-nematic phase transition}
In this Subsection, we will assess the accuracy of the various DFT approximations introduced in Section 
\ref{sec:DFT} to investigate the N-N$_{S}^{*}$ transition, through a direct comparison with numerical simulations.
To this aim, we will consider the particular case of helices having full translational and azimuthal freedom, but with their $\hat{\mathbf{u}}$ axis parallel to the $\hat{\mathbf{n}}$ director.  This assumption, partly justified by the observation that the N-N$^*_{S}$ phase transition occurs at 
large values of the nematic order parameter, has the advantage of simplifying the theoretical treatment and considerably reducing its computational cost. This notwithstanding, it can still be useful for several different reasons. Firstly, it provides direct insights into the order of the N-N$_{S}^{*}$ transition, and its relationship with the helix morphology,  
decoupled from the effects of the particle structure on the stability of the nematic phase. Secondly, it allows us to probe the reliability of theory to describe this rich and unconventional scenario. Finally, it is an interesting problem on its own right as the behavior of non-convex hard particles has so far largely overlooked
in spite of the large number of examples in real systems.

\begin{figure}
\includegraphics[width=8.0cm]{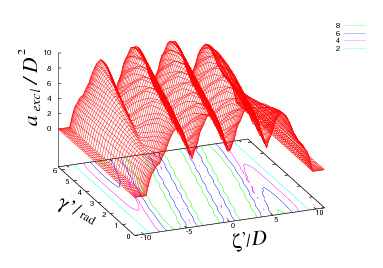}
\caption{
The function $a_{excl}(\zeta^{\prime}, \gamma^{\prime})$ (see Eq. \ref{ddue}), calculated for helices with $r=0.4$ and $p=4.322$. 
}
\label{exclareaexamp}
\end{figure}

We remark that, unlike previous cases, we have here used the number of complete turns ($n$) as input variable, thus resulting
in a non-integer pitch value $p$. The relation between $n$ and $p$ can be found in Ref.\cite{Frezza13}.
Fig. \ref{exclareaexamp} gives an example of the function $a_{excl}(\zeta^{\prime}, \gamma^{\prime})$  (Eq. \ref{ddue}), for 
the case with $r=0.4$ and $p=4.322$.  
This function exhibits oscillations, whose number reflects the number of turns in the helix.  
Oscillations are comprised between the values of excluded area for cylinders enclosing 
the whole helix ($a_{excl}=\pi (2r+1)^2$) and for cylinders enclosing a linear 
chain of beads
($a_{excl}=\pi$), as they should. 
The decrease of  $a_{excl}$ resulting from the interpenetration of helices is related to 
the entropy gain that drives the formation of the N$_{S}^*$ phase.
\begin{figure}
\includegraphics[width=8.0cm]{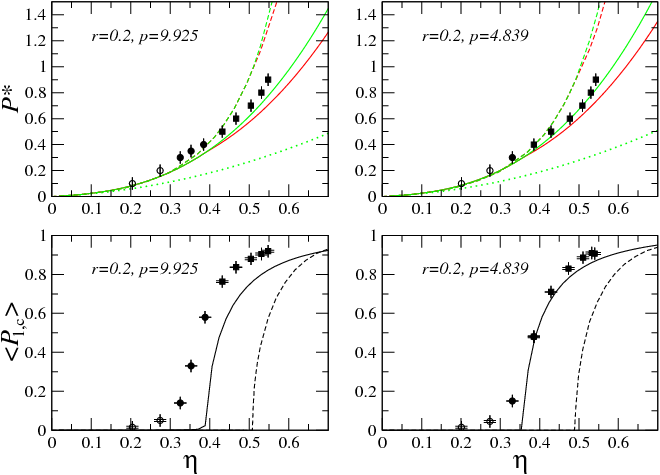}
\caption{Equation of state (top) and 
screw-nematic order parameter as a function of the volume fraction $\eta$ (bottom),
for perfectly aligned hard helices with 
$r=0.2$ and $p=9.925$ (left) and $p=4.839$ (right).
Results from Onsager theory (dotted), 
from modified PL theory (dashed) and from third-virial theory (solid).
Top panels: green is for the N phase and red for the N$_{S}^*$ phase. 
Symbols are results from MC simulations: N (empty circles), N$^*_{S}$ 
(full circles) and Sm$_{S}^*$ (full squares).}
\label{r0.2n1}
\end{figure}
\begin{figure}
\includegraphics[width=8.0cm]{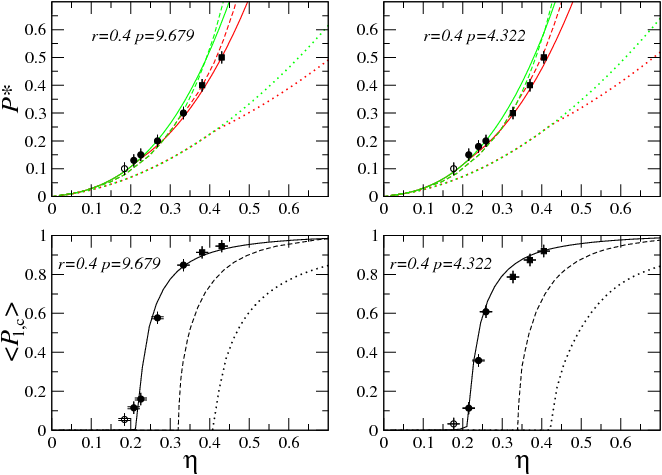}
\caption{Equation of state (top) and 
screw-nematic order parameter as a function of the volume fraction $\eta$ (bottom),
for perfectly aligned hard helices 
with $r=0.4$ and $p=9.679$ (left) and $p=4.322$ (right).
Results from Onsager theory (dotted), 
from modified PL theory (dashed) and from third-virial theory (solid).
Top panels: green is for the N phase and red for the N$_{S}^*$ phase. 
Symbols are results from MC simulations: N (empty circles), N$^*_{S}$ 
(full circles) and Sm$_{S}^*$ (full squares).}
\label{r0.4n1}
\end{figure}

Fig. \ref{r0.2n1} shows equation of state and screw-nematic order parameter as
a function of $\eta$ for the cases with $r=0.2$ and $p=9.925$ and $4.839$,
while Fig. \ref{r0.4n1} shows the same quantities for the cases with
$r=0.4$ and $p=9.679$ and $4.322$.
These figures provide results from second- and third-virial theories along with
corresponding MC simulation data.
We can see that for these helices, which have a pitch larger than the bead diameter,
the location of the phase transition is essentially determined by the radius $r$,
depending mildly on $p$, and
in particular it occurs at increasing density with decreasing radius.
This can be understood considering that 
less curly helices have weaker oscillations of $a_{excl}$, 
thus a lower entropy gain is achieved for them upon the settling in of the screw-like order.
Present results would seem to hardly reconcile with the phase diagrams shown in Sec. III A,
where the  N-N$^*_{S}$ phase transition occurs at a volume fraction that increases on
moving from $r=0.2$, $p=4$, to $r=0.4$, $p=4$ and then to $r=0.2$, $p=8$.
However, it has to be recalled that, on one hand, 
the hard helices considered here have 
the same contour length and their effective aspect ratio thus decreases 
on going from straight to curly particles and, 
on the other hand, that in the MC-NPT simulations 
 helices are freely rotating.
The onset of any liquid-crystalline order thus always competes with the I phase,
favoured at low  densities and 
whose stability shifts to higher densities as 
the effective aspect ratio becomes smaller.

Compared to MC data for perfectly aligned helices, while a purely second-virial theory alone proves overall inadequate,
significant improvements are achieved including PL correction and the third-virial term.
By looking at each and confronting one another all these figures,
it seems that
the predictions of a third-virial theory improve as $r$ increases and $p$ decreases,
to such an extent that, for the cases with $r$=0.4
quantitative agreement is there between theory and simulations.
This situation is in a way spoilt by the fact that
the phase observed in simulations at higher densities is actually
a Sm$_{S}^*$ \cite{note6}  rather than a N$_{S}^*$ phase.
Theoretical calculations that include the former are not available at present
(it would amount to deal with Eq.\ref{elibsme} rather than simpler Eq.\ref{due}).
In spite of this caveat, results from a third-virial theory are considered overall encouraging.
\section{Discussion}
\label{subsec:physical}
We are now in the position to try and understand  the physical origin 
of the phase sequence exhibited by helical particles.  
The nematic phase spans a density range that can be subdivided in two regions, 
the first being the conventional N phase at lower densities, 
the second being the screw-like N$_{S}^*$ phase at higher density.
The relative width of the two regions varies depending upon the helix parameters, 
but with the screw-like N$_{S}^{*}$ phase always popping out at the right edge of the nematic window, 
while the  N phase may or may not be present. 
Indeed, the latter can be absent altogether for sufficiently high degree of curliness
as shown by our results in the case of $r=0.4$ and $p=4$. 
Indeed, on increasing  the helical twist, the relative width of the screw-like region increases with respect to 
the conventional N region, until the latter eventually disappears. The underlying mechanism is as follows.

\begin{figure}
\begin{center}
\includegraphics[width=4.0cm]{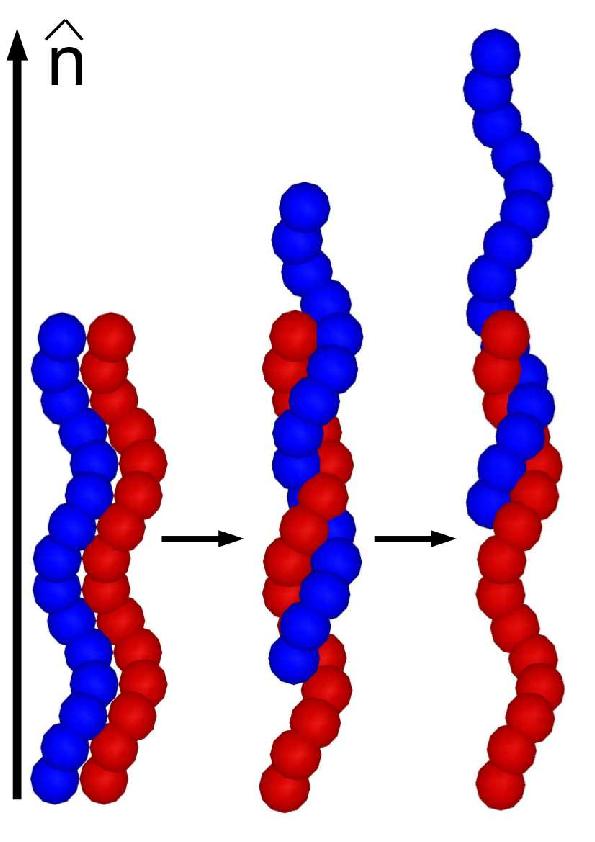} 
\end{center}
\caption{The screw-like coupled translation and rotation.}
\label{fig:screw-like}
\end{figure}

Imagine to have two neighbouring helices that are sufficiently far apart 
to be able to rotate about their own main axis.
Effectively, they  behave as cylinders, 
their specific helix character being rather irrelevant and 
the liquid-crystal phase they may be in is the conventional N phase. 
This is however no longer the case if the two helices are in close contact one another so that
neighbouring grooves significantly intrude into each other into a in-phase locked configuration, 
as illustrated in Fig.\ref{fig:screw-like}. 
Because of this azimuthal locking of the C$_2$ axes,
there is then a severe limitation of the rotational entropy and 
there must then be a correspondingly higher gain in translational entropy 
in order for the new chiral nematic phase to be stable with respect to the N phase. 
This is achieved through a screw-like organization,
schematically illustrated in Fig.\ref{fig:screw-like} 
where the right helix rotates about its own axis by performing an additional translation along the same axis.  
\begin{figure}
\includegraphics[width=6.2cm]{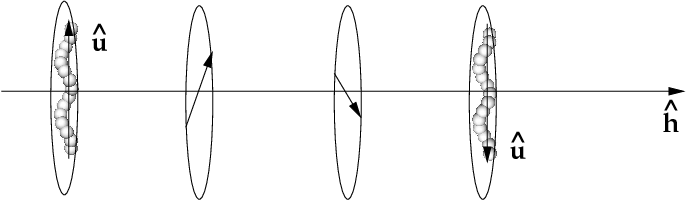} \\
\includegraphics[width=6.2cm]{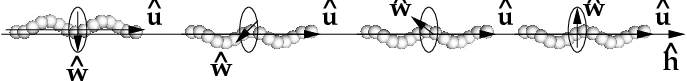} \\
\includegraphics[width=6.2cm]{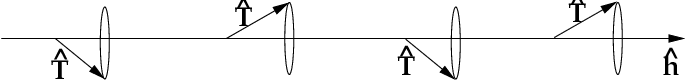} 
\caption{Cartoon of the cholesteric (top), and of the screw-nematic (middle) organization. Here $\widehat{\mathbf{h}}$ is a unit vector parallel to the axis around which
either the  molecular axis $\widehat{\mathbf{u}}$ (cholesteric) or $\widehat{\mathbf{w}}$ (screw-nematic) spirals, respectively.  In the latter case, the tip of the tangent
to the helices $\widehat{\mathbf{T}}$ forms a conical path (bottom).
}
\label{fig:conical_vs_screw}
\end{figure}
 The N$_{S}^{*}$  is a novel kind of chiral nematic phase, different from the well known cholesteric. The latter may be formed in general by any kind of chiral mesogenic particles, whereas the former is special of helical particles. It may be useful to highlight analogies and differences between  these phases. 
In the cholesteric phase the $\widehat{\mathbf{u}}$ axes of helices  
spiral around a perpendicular axis ($\widehat{\mathbf{h}}$) as illustrated in Fig.\ref{fig:conical_vs_screw} (top). The order of the $\widehat{\mathbf{u}}$ axes  is non-polar, i.e. there is up-down symmetry.  
In the N$_{S}^{*}$ phase 
the $\widehat{\mathbf{u}}$ axes of helices are preferentially aligned along the same direction throughout the sample, but the transversal $\widehat{\mathbf{w}}$ axes spiral around this direction, as depicted in Fig. \ref{fig:conical_vs_screw} (center). In this case the $\widehat{\mathbf{w}}$ axes have polar order, i.e. they preferentially point in the same direction.   
Another important difference between screw-nematic and cholesteric is the length scale of the phase periodicity, which is equal to the pitch of the helical particles in the former, and orders of magnitude longer in the latter. This is the reason why the screw-like organization, unlike the cholesteric, can be observed in simulations with box sizes of a few molecular lengths and standard periodic boundary conditions.

A different description of the screw-nematic phase could be made, in terms of the
Frenet frame routinely used for the description of flexible and semi-flexible polymers 
(e.g. \cite{Kamien02}).
This however provides a redundant description for rigid objects of finite lengths as the helices considered in the present work. 
Using this picture, the tip of the local tangent the helices ( $\widehat{\mathbf{T}}$) follows a conical path on moving along the director $\widehat{\mathbf{n}}$, 
due to the variation of the azimuthal angle at fixed polar angle, 
as shown in Fig.\ref{fig:conical_vs_screw} (bottom). 
This was the description used in Ref.\cite{Barry06} to explain 
the experimental results for helical flagella and that led
to denoting this phase as conical.\cite{Meyer68,Kamien96}
We also have found it useful for the visualization of snapshots 
colour coded according to the local tangent of helices. 

The situation is expectedly much more complex in smectic phases, 
where screw-like organization, layering  and hexatic order 
may compete and combine one another. 
As the system is entering a smectic phase, 
there still exists a non-negligible fraction of interlayer helices 
while positional ordering along the $\widehat{\mathbf{n}}$ direction progressively increases. 
These helices lying in the interlayer regions provide a bridge between two adjacent layers, 
and allow a screw-like organization to be present througout the whole smectic phase.  
When  the concentration of helices is still moderate,
hexatic ordering is not significantly present  
and the first  smectic phase encountered upon increasing density  is the Sm$_{A,S}^{*}$.
As in the nematic phase, this roto-translational coupling may be more or less effective depending upon the morphology of the helix, but is always present in the initial part of the smectic phases.

As pressure is further increased, the hexatic order gradually sets in,
typically accompanied by a concomitant increase of 
the azimuthal correlation of the $\widehat{\mathbf{w}}$ axes of 
helices within each layer. 
This may lead to two different, and up to a certain extent competing, effects.
The first possibility is that helices are azimuthally aligned within each each layer, 
but with neither positional nor orientational correlations between layers. 
Each layer can also in principle rigidly slide with respect to next layers, 
to gain translational entropy. 
Under these conditions, layering is very strong as testified by 
the solid-like peaks in the observed $g_{\|} (R_{\|})$.
This situation, depicted in the cartoon of Fig.\ref{fig:stratiA_vs_stratiB} (right panel),
differs from the conventional smectic B phase for the presence of the in-plane (polar) correlation between the $\widehat{\mathbf{w}}$ axes of helices, and hence the phase was denoted 
as Sm$_{B,p}$. 

The alternative scenario stems from the possibility 
that tips of helices protruding out from a layer are still able to propagate the ordering to the
neighboring layers. 
Thus, helices belonging to different layers stack on top of each other along $\widehat{\mathbf{n}}$ 
to form parallel, "infinitely" long, helices. 
The alignement of the the $\widehat{\mathbf{w}}$  
axes then translates into a screw-like ordering propagating across the layers.
Under these conditions, clearly different layers are strongly correlated with each other.
The layer structure along $\widehat{\mathbf{n}}$ is preserved, but 
positional ordering along $\widehat{\mathbf{n}}$ is 
less effective, due to the presence of protruding helices, 
as testified by the reduced peaked structure of $g_{\|} (R_{\|})$. 
This is the situation represented in the left panel of Fig.\ref{fig:stratiA_vs_stratiB}. 
We denoted this phase as Sm$_{B,S}^{*}$, 
because it couples layering with hexatic positional order and screw-like azimuthal correlations. 
The entropic advantage of this scheme is to form a set of "infinite" parallel helices,
which allows the favourable screw-like motion to be still operative. 
One may envisage the additional presence of a columnar phase in the case of helices with sufficiently long contour lengths, well
beyond those considered here. This is a subject that deserves a dedicated study.
\section{Conclusions and outlook}
\label{sec:conclusions}

\begin{table}[h]
\small
\begin{tabular*}{0.5\textwidth}{@{\extracolsep{\fill}}lll}
\hline
Phase & Code & Organisation  type     \\
\hline
Conventional nematic       & N                 & $(a)$  \\
Screw- nematic             & N$_{S}^{*}$        & $(b)$  \\
Screw-smectic A            & Sm$_{A,S}^{*}$     & $(c)$   \\
Polar smectic B            & Sm$_{B,p}$         & $(d)$   \\   
Screw-smectic B            & Sm$_{B,S}^{*}$     &  $(e)$  \\
\hline 
\end{tabular*}
\caption{Summary of the different phases exhibited by hard helices. 
$(a)$$\widehat{\mathbf{u}}$ axes oriented along  $\widehat{\mathbf{n}}$;
$(b)$ as in N with azimuthal coupling of $\widehat{\mathbf{w}}$ axes along  $\widehat{\mathbf{c}}$  that in turn spirals about $\widehat{\mathbf{n}}$; $(c)$as in Sm$_{A}$ with azimuthal coupling of $\widehat{\mathbf{w}}$ axes along $\widehat{\mathbf{c}}$ that in turn spirals about $\widehat{\mathbf{n}}$; $(d)$ as in Sm$_{A}$ with additional in-plane polar and hexatic order; $(e)$ as in Sm$_{A,S}^{*}$ with additional hexatic order.
} 
\label{tab:1}
\end{table}
In this work, we have studied the self-assembly properties of systems of hard helices as 
a function of helix morphology.
Helical particles have been modelled as a set of fused hard spheres properly arranged 
to form a rigid helix of a fixed contour length. 
Using a combination of numerical simulations and density functional theories, 
we have analyzed the sequence of different liquid crystal phases appearing at increasing density, using
a set of suitable order parameters and correlation functions. 

The rich and unconventional polymorphism that we found is in striking contrast 
with the conventional wisdom of assimilating the phase behaviour of helical particles to that of rods, 
an assumption commonly adopted also in the analysis of experiments on helical (bio)polymers.
Table \ref{tab:1}  summarizes the distinctive features of all phases discussed in this work. 

The first novel phase encountered with increasing density is the 
screw-nematic (N$_{S}^{*}$). As neighboring helices tend to lock into an in-phase nematic configuration by 
an azimuthal correlation of the helix 
$\widehat{\mathbf{w}}$ axes along 
a common direction $\widehat{\mathbf{c}}$, 
there must be a concomitant gain in translational entropy counterbalancing 
that loss of rotational entropy for this new phase to be stable. 
This is achieved through a translational-rotational coupling
where $\widehat{\mathbf{c}}$ spirals around the main nematic director $\widehat{\mathbf{n}}$, 
with a periodicity equal to the pitch of the single helix. 
We have also implemented a density functional theory with increasing degrees of accuracy, for the screw-nematic N$_{S}^{*}$ phase, under the assumption
of perfectly aligned helices, and tested its accuracy with numerical simulations on the same system. We find the results of the most accurate versions of the
theory in reasonably good quantitative agreement with numerical simulations. 

With increasing density a smectic A phase with screw-like order (Sm$_{A,S}^{*}$) can appear, which differs from the N$_{S}^{*}$ for the presence of layers. However helices laying in the interlayer regions provide a bridge between adjacent layers, 
which allows to keep the  screw-like organization.  
As density increases, positional ordering along $\widehat{\mathbf{n}}$ also increases, 
while in-plane hexatic  order tends to set in. 
This leads to the formation of either a polar Sm$_{B,p}$ phase, 
characterized by the fact that different layers can rotate and translate independently of each other with no coupling between orientations of $\widehat{\mathbf{w}}$ 
axes in different layers,  or of 
a Sm$_{B,S}^{*}$, with screw-like coupling between adjacent layers, 
Our results indicate that a 
Sm$_{B,p}$ phase is more favoured for slender helices, 
with a gradual transition to screw-smectic  Sm$_{B,S}^{*}$ phases for curlier particles. 
At even higher densities, a very compact phase, that we generally labelled as C, is achieved. 
This phase is likely to display some regular crystal structure,
as indicated by the regular peaks in several correlation functions the we have monitored. 
A detailed study of this phase will be discussed elsewhere. 

The results presented here call for experimental verification. To this purpose, an important distinctive feature of most of the novel phases identified in our study is the presence of a phase modulation with the periodicity equal to the pitch of the constituting helical particles. In principle, helical biopolymers such as DNA or helical colloidal particles, appear as good candidates for this investigation. \cite{Nakata07,Zanchetta10,DeMichele12}
Indeed, a screw-nematic phase was already observed a few years ago in
colloidal suspensions of helical flagella 
isolated from prokaryotic bacteria.\cite{Barry06} 
The helical pitch $p$ of these particles is $\mu$m in size and hence the phase modulation is easily visible under polarized optical  microscopy. However, for chiral polymers  typical values of the pitch are in the nm range, which is far too small to be observable by any optical microscopy. In this case the experimental determination of the phase periodicity  constitutes an experimental challenge. We hope that our study can stimulate new work in this direction.

\begin{acknowledgments}
H.B.K., E.F. A.F. and A.G. gratefully acknowledge support from PRIN-MIUR 2010-2011 project (contract 2010LKE4CC). 
G.C. is grateful to the Government of Spain for the award of a Ram\'{o}n y Cajal research fellowship. 
H.B.K., A.G. and T.S.H. also acknowledge the support of a Cooperlink bilateral agreement Italy-Australia.
\end{acknowledgments}


\begin{thebibliography}{99}
\bibitem{note3}{The effective length of helices is taken equal to $\Lambda$+$D$, where $\Lambda$ is the Euclidean length \cite{Frezza13} to which  the bead diameter is added.}
\bibitem{note6} {We did not pursue the further characterisation of this screw-smectic phase      
as this would have been contingent here.}
\end{thebibliography}
\end{document}